\documentclass[12pt]{article}

\usepackage{amsfonts}
\usepackage{amssymb}
\usepackage{amsmath}
\usepackage{latexsym}

\def\R{{\cal R}}

\def\beq{\begin{equation}}
\def\eeq{\end{equation}}

\def\bq{\begin{quote}}
\def\eq{\end{quote}}

\newcommand{\non}{\nonumber}
\newcommand{\be}{\begin{equation}}
\newcommand{\ee}{\end{equation}}
\newcommand{\bea}{\begin{eqnarray}}
\newcommand{\eea}{\end{eqnarray}}
\newcommand{\ba}{\begin{array}}
\newcommand{\ea}{\end{array}}
\newcommand{\al}{\alpha}

\newcommand{\pa}{\partial}
\newcommand{\ep}{\epsilon}

\newcommand{\ta}{\tau}
\newcommand{\tta}{\tilde\tau}

\newcommand{\om}{\omega}
\newcommand{\Om}{\Omega}

\newcommand{\De}{\Delta}
\newcommand{\vphi}{\varphi}

\newcommand{\rar}{\rightarrow}

\allowdisplaybreaks
\newcounter{mycount}

\begin{document}

\begin{titlepage}
\vspace{-1mm}
\vspace{1mm}
\begin{center}{\bf\Large\sf Solvability of the Hamiltonians related
to exceptional root spaces: rational case}
\end{center}

\begin{center}
{\bf Konstantin G.~Boreskov}{\normalsize
\footnote{boreskov@heron.itep.ru},\\ {\em Institute for
Theoretical and Experimental Physics, \\ Moscow 117259, Russia}\\
\vspace{2mm}}

{\bf Alexander V.~Turbiner{\normalsize \footnote{
turbiner@nuclecu.unam.mx}${}^{,} $\footnote{On leave of absence
from the Institute for Theoretical and Experimental Physics, \\
\indent \hspace{5pt} Moscow 117259, Russia.}}}
\\[2mm]
and
\\[2mm]
{\bf Juan Carlos Lopez Vieyra {\normalsize
\footnote{vieyra@nuclecu.unam.mx}}},
\\ {\em
Instituto de Ciencias Nucleares, UNAM, A.P. 70-543,\\ 04510 Mexico
D.F., Mexico} \\ \vspace{2mm}

{\bf\large Abstract}
\end{center}
\small{
\begin{quote}
  Solvability of the rational quantum integrable systems related to
  exceptional root spaces $G_2, F_4$ is re-examined and for
  $E_{6,7,8}$ is established in the framework of a unified approach.
  It is shown the Hamiltonians take algebraic form being written in a
  certain Weyl-invariant variables. It is demonstrated that for each
  Hamiltonian the finite-dimensional invariant subspaces are made from
  polynomials and they form an infinite flag. A notion of minimal flag
  is introduced and minimal flag for each Hamiltonian is found.
  Corresponding eigenvalues are calculated explicitly while the
  eigenfunctions can be computed by pure linear algebra means for {\it
    arbitrary} values of the coupling constants. The Hamiltonian of
  each model can be expressed in the algebraic form as a second degree
  polynomial in the generators of some infinite-dimensional but
  finitely-generated Lie algebra of differential operators, taken in a
  finite-dimensional representation.
\end{quote}}
\vskip .5cm
\end{titlepage}
\vskip 2mm

\setcounter{equation}{0}

\section{Introduction}

Up to now the Hamiltonian Reduction Method which also is called the
Projection Method \cite{Olshanetsky:1977,Olshanetsky:1983} provides a
unique opportunity to construct non-trivial multidimensional
completely integrable quantum (and classical) Hamiltonians. These
Hamiltonians are associated with root systems, they are related with
the Laplace-Beltrami operators on symmetric spaces. Their rational
(trigonometric) versions have remarkable properties: (i) their
eigenvalues are known explicitly being at most the second degree
polynomial in quantum numbers, (ii) any eigenfunction has a form of
the ground state eigenfunction multiplied by a polynomial in the
Cartesian (or exponential in Cartesian) coordinates. These two
specific types of the Hamiltonians appear naturally in this approach
-- with rational and trigonometric potentials, correspondingly,
characterized by the second-order poles in potentials on the
boundaries of the configuration space.  However, soon after a
discovery of these Hamiltonians it became clear
\cite{Olshanetsky:1983} there exists a straightforward generalization
of these Hamiltonians to the case of arbitrary coupling constants
without breaking any nice property. It led to a loss of immediate
group theoretical interpretation. However, a property of solvability
remained to hold. It gave a hint on existence of a more general
formalism where above-mentioned Hamiltonians with arbitrary coupling
constants appear naturally. An idea was to connect solvability with
possible existence of an intrinsic hidden algebraic structure
\cite{Turbiner:1994, Ruhl:1995, Brink:1997}. It turns out to be true.
For arbitrary coupling constants these Hamiltonians admit an algebraic
form (where polynomial coefficient functions occur in front of
derivatives) and they are related with elements of the universal
enveloping algebra of some algebras of differential operators acting
in the space of invariants of the corresponding root space. Such an
algebra was called the {\it hidden} algebra of the Hamiltonian. It was
found that for all $A_n, B_n, C_n, D_n, BC_n$ rational and
trigonometric models this algebra is the {\it same} (!) -- it is the
maximal affine subalgebra of the $gl_n$-algebra realized by the first
order differential operators in ${\bf R^n}$ and taken in symmetric
representation \cite{Ruhl:1995, Brink:1997}. Thus, one can state that
all these models are nothing but different appearances of a {\it
  single} model characterized by the hidden algebra $gl_n$. Similar
situation holds for the SUSY generalizations of above models - all of
them turned out to be associated to the hidden superalgebra
$gl(n|n-1)$, see \cite{Brink:1997}).

However, one can naturally expect that the situation is drastically
different for the Hamiltonians related to the exceptional algebras -
each Hamiltonian is characterized by its own hidden algebra which is
different for different Hamiltonians.  A first indication stemmed from
a study of the $G_2$ rational and trigonometric models, where the both
models were characterized by the same hidden algebra, but this algebra
turned out to be a certain infinite-dimensional, finitely-generated
algebra of the differential operators $g^{(2)}$, which is a subalgebra
of the algebra of the differential operators on the real plane,
$g^{(2)} \subset \mbox{diff}(2,\mathbb{R})$ \cite{Rosenbaum:1998,
  Turbiner:1998}.  Later, it was shown that the similar situation
holds for the rational and trigonometric $F_4$ models \cite{blt}: both
models possess the same hidden algebra of the differential operators
$f^{(4)}$, which is a subalgebra of the algebra of the differential
operators on $\mathbb{R}^4$,  $f^{(4)} \subset \mbox{diff}(4,\mathbb{R})$.

For all previous studies a crucial point was to find a set of
variables in which the Hamiltonian under investigation takes an
algebraic form and reveal their meaning. A simple observation made in
\cite{Ruhl:1995} was to consider the variables which incorporate all
symmetries of the problem in hand. In particular, these symmetries
contain (or sometimes even coincide with) the Weyl group of the
associated root system. A natural idea is to take invariants of the
fixed degrees of the Weyl group as variables. Many years ago
V.I.~Arnold \cite{Arnold:1976} pointed out that the contravariant flat
metric on the space of orbits of any Coxeter group written in terms of
the polynomial invariants has polynomial matrix elements. It implies
that the coefficient functions in front of the second derivatives in
the Laplace-Beltrami operator are polynomials in invariants of the
Weyl group. This result was rediscovered (and then generalized) later
in \cite{Ruhl:1995}, \cite{Brink:1997}, \cite{Rosenbaum:1998} and
\cite{blt} for $A_n$, $BC_n$, $G_2$ and $F_4$ algebras, respectively.
It was shown that the algebraic structure persists for the whole
Laplace-Beltrami operator: the coefficient functions in front of the
first derivatives are polynomials as well. Further generalization was
that similar statement is valid for the entire set of the rational
Hamiltonians which are a combination of Laplace-Beltrami operator and
a potential -- after the gauge (similarity) transformation with the
ground state eigenfunction the coefficient functions in front of the
first derivatives remain polynomials\footnote{ Similar results we also
  obtained in above-mentioned works for the trigonometric $A_n$,
  $BC_n$, $G_2$ and $F_4$ Hamiltonians when the trigonometric Weyl
  invariants (which mean the Weyl invariants periodic in each
  variable) are used as coordinates.  Even for elliptic $A_1$ and
  $BC_n$ models the elliptic Weyl invariants allow to get the
  polynomials coefficients in front of derivatives (see
  \cite{Turbiner:1989} and \cite{Gomez-U:2000} for details) }.  In the
present paper a certain universal prescription about a choice of
coordinates is given.

Recently, it was found that the existence of algebraic forms and
knowledge of hidden algebraic structure of the above-mentioned
Hamiltonians allows to consider their perturbations in a constructive
way. One can develop an {\it algebraic} perturbation theory, where all
corrections can be obtained by pure linear algebra means
\cite{Turbiner:1999, Turbiner:2002, Turbiner:2003}.

The goal of the present article is to carry out a study of the
solvability of the rational models related to the exceptional root
spaces. We introduce a general formalism which allows to study all
these models on equal footing (Section 2). In Sections 3, 4 the
solvability of the $G_2$ and $F_4$ models is re-examined in a new
formalism, while in the Section 5--7 the solvability is established
for $E_{6,7,8}$ rational models. In Conclusion the results are
summarized. Almost all results were obtained with the help of MAPLE 7
and MAPLE 8 programs together with the package COXETER created by
J.~Stembridge.

\setcounter{equation}{0}

\section{Generalities}

Let us consider a quantum system described by rational
Hamiltonian associated with a root system $\R$ of algebra $g$ of
rank $N$:
\begin{align}
\label{Ham_gen}
 {\cal H} = \frac{1}{2}\sum_{k=1}^{N}
 \bigg[-\frac{\pa^{2}}{\pa x_{k}^{2}}+ \om^2 x_k^2 \bigg]\ +
 \frac{1}{2}\sum_{\alpha\in \R_{+}}
 g_{|\alpha|}|\,\alpha|^{\,2}\frac{1}{(\alpha\cdot x)^2} ~,
\end{align}
where $\al \in \R_+$ are positive roots of the system $\R$ which are
vectors in $\mathbb{R}^N$, $x={(x_1,\ldots,x_N)}$ is a set Cartesian
coordinates, $|\alpha|^2=\sum_1^N \alpha_k^2$, and the scalar product
$(\alpha\cdot x)=\sum_1^N \alpha_k x_k$, $\om$ is a parameter.
Coupling constants $g_{|\alpha|}$ are assumed to be equal for roots of
the same length. Hence for the $A_n$ case there is a single coupling
constant, for the $BC_n$ case there are three coupling constants, etc.
For some algebras ($G_2$, $E_6$, $E_7$) it is convenient to embed the
roots into the vector space of higher dimension $\mathbb{R}^{N+n}$
($n=1$ or $2$). In general, the Hamiltonians of this type describe a
quantum particle in multidimensional space, although for $A_n$ and
$G_2$ cases these Hamiltonians allow another interpretation as the
Hamiltonians describing many-body systems, while for $B_n,C_n,D_n$
they correspond to many-body systems on the space with mirror .

Consider the spectral problem for the Hamiltonian ${\cal H}$,
\begin{equation}
\label{1.1}
    {\cal H} \Psi (x)\ =\ E \Psi (x)\ ,
\end{equation}
where the configuration space is the Weyl chamber and $\Psi$ should be
from the corresponding Hilbert space. Let us make a gauge rotation of
the Hamiltonian taking the ground state eigenfunction $\Psi_0$ as a
gauge factor
\begin{equation}
\label{e1.2}
 h = - 2(\Psi_0(x))^{-1} ({\cal H} - E_0) \Psi_0(x) \ ,
\end{equation}
where $E_0$ is the ground state energy of the Hamiltonian
(\ref{Ham_gen}) (see e.g. \cite{Olshanetsky:1983}),
\begin{align}
\label{e1.2a}
%
 E_0 = \om \bigg(\frac{N}{2} + \sum_{\al\in \R_+}g_{|\alpha|}\bigg) ~.
\end{align}
It should be mentioned that for the $A_n$ and $G_2$ cases the energy
$E_0$ does include the ground state energy of center-of-mass motion.

The ground state eigenfunction of (\ref{Ham_gen}) has a form
\begin{equation}
\label{gr-st-ex}
 \Psi_0\ =\ \De_{g} \exp{\big(-\frac{\om}{2} t_2^{(\Om)} \big)}\ ,
\end{equation}
where
\[
\De_{g}\ =\ \prod_{\R_{+}} (\al_k, y)^{\nu_{|\al|}}
\]
and $\nu_{|\al|}$ are defined through
$g_{|\al}=\nu_{|\al|}(\nu_{|\al|}-1)$ and assumed to be equal for
roots of the same length. For example, if all roots are of the same
length as for $A_n$ case, we put $\nu_{|\al|}= \nu$. In the case of
roots of two different lenghts as for $G_2$ case we denote
$\nu_{|\al|}= \nu$ for the short roots and $\nu_{|\al|}= \mu$ for the
long roots. $t_2^{(\Om)}$ is the invariant of the degree two (for
definition see below).

A new spectral problem arises
\begin{equation}
\label{1.3}
    h \vphi (x)\ =\ -\ep \vphi (x)\ ,
\end{equation}
with a new spectral parameter $\ep=2(E-E_0)$. If in (\ref{1.1}) the
boundary condition means normalizability of the eigenfunction $\Psi
(x)$, then for (\ref{1.3}) it requires the normalizability of $\vphi
(x)$ with the weight factor $\Psi_0^2 (x)$.  By construction, the
lowest eigenvalue $\ep_0=0$ and the lowest eigenfunction is
$\vphi_0=\mbox{const}$. Our goal is to find, by a change of variables,
an {\it algebraic} form of the operator $h$ (if it exists).

{\bf Definition:}  a linear differential operator with
polynomial coefficients is called  {\em algebraic}.

In order to find these new variables we assume that they respect the
symmetries of the Hamiltonian \cite{Ruhl:1995}. In our case it means
the invariance under the group of automorphisms $A_g$ of the $g$ root
space. This group includes or coincides with the Weyl group $W_g$. The
algebraically independent invariant polynomials of the {\it lowest}
possible degrees $a$ generate the algebra $S^{A_g}$ of $A_g$-invariant
polynomials. The powers $a$ are the {\em degrees} of the group $W_g$.
A particular form of these polynomials (denoted below as
$t_a^{(\Om)}$) can be found by averaging elementary monomials $(\om
\cdot x)^{a}$ over some group orbit $\Om$,
\begin{align}
\label{orbit}
 t_{a}^{(\Om)}(x) = \sum_{\om\in\Om} (\om \cdot x)^{a}\ ,
\end{align}
(see e.g. \cite{Bourbaki}), where $x$'s are some formal variables.
There exists a certain ambiguity in connecting these variables with
the variables appearing in the Hamiltonian under study.  It is worth
mentioning that for any Lie-algebra $g$ there exists a second degree
invariant $t_{2}^{(\Om)}(x)$ which does not depend on chosen orbit.
It is namely this invariant which defines the exponent in the ground
state eigenfunction (see above).  One of natural connections is to
identify $x$'s with Cartesian coordinates. Later on we will use the
expressions (\ref{orbit}) as new variables in Hamiltonians under study
and will call them the {\it orbit variables}.

In general, making averaging over different orbits in
(\ref{orbit}) we obtain algebraically related invariants, except
for a case when averaging over a specific orbit leads to vanishing
results for certain invariant(s). Without  loss of generality,
these orbits can be discarded. In fact, the variables which were
successfully used to solve the cases $A_n$ and $BC_n$ in
\cite{Ruhl:1995} and \cite{Brink:1997}, correspondingly, as well
as for $G_2$ and $F_4$ (see \cite{Rosenbaum:1998} and \cite{blt},
correspondingly) can be easily obtained through the formulas
(\ref{orbit}).

The invariants of the fixed degrees are defined up to a polynomial in
invariants of the lower degrees. This ambiguity plays an important
role in obtaining the algebraic forms of Hamiltonians -- these forms
depend on the choice of variables, special combinations of variables
only result in the simplest form which correspond to preservation of
the minimal flag (see below).

Now we introduce a notion of {\it exact-solvability}. Let us assume
that the operator $h$ possesses infinitely-many finite-dimensional
invariant subspaces ${\cal V}_n,\quad n=0,1\ldots$, which can be
ordered forming an infinite flag
\[
{\cal V}_0 \subset  {\cal V}_1 \subset {\cal V}_2 \subset \ldots
 \subset  {\cal V}_n  \subset \ldots \ ,
\]
(or, filtration)  ${\cal V}$.  Therefore one can say that the
operator $h$ preserves the flag ${\cal V}$.\\

{\bf Definition } \cite{Turbiner:1994}

\begin{itemize}
\item An operator $h$ which preserves an infinite flag of explicitly
  defined finite-dimensional spaces ${\cal V}$ is called {\it
    exactly-solvable operator with flag ${\cal V}$}. We assume that
  the flag ${\cal V}$ is {\it dense}: between any two subsequent
  spaces there is no space of intermediate dimension which might
  belong to the flag.

\item If given $h$ preserves several flags and among them
there is a flag for which $\mbox{dim} {\cal V}_n$ is {\it maximal}
for any given $n$, such a flag is called {\it minimal}.
\end{itemize}

Below we will deal with certain linear spaces of polynomials in
several variables.\\

{\bf Definition}


Consider the triangular linear space of polynomials in $k$
variables
\begin{equation}
\label{flag}
 {\cal P}_{n}^{(f_1, \ldots, f_k)} \ = \ \langle s_1^{p_1}
s_2^{p_2} \ldots s_k^{p_k} | 0 \leq f_1 p_1 + f_2 p_2 +\ldots
+ f_k p_k \leq n \rangle\ ,
\end{equation}
where $f$'s are positive integer numbers and $n$ is integer.  {\it
  Characteristic vector} is a vector with components which are equal
to the coefficients (weights) $f_i$ in front of $p_i$:
\begin{equation}
\label{char.vec.}
 \vec f = (f_1, f_2, \ldots f_k)\ .
\end{equation}
In other words, $\vec f$ defines an action of $\mathbb C^*$ on the
space $\mathbb C[s_1,\ldots ,s_k]$ of polynomials in $k$ variables. The
flag is defined using the induced grading.  

The characteristic vector is defined up to a multiplicative integer
factor which we choose to be minimal. In most of the examples $f_1=1$.
Taking a sequence of the spaces characterized by growing integer
numbers $n$ we arrive at a flag which has ${\cal P}_{n}^{(f_1, \ldots,
  f_k)}$ as {\it generating} linear space. In the examples $n$ takes
consecutive integer values, $n=0,1,2,\ldots$. We call such a flag
${\cal P}^{(f_1, \ldots, f_k)}$.

All Hamiltonians we are going to study are of quite special type.  All
their flags of invariant subspaces, that we were able to find so far,
are the flags of polynomials of the form (\ref{flag}). Among these
flags there always exists a minimal flag of a special form --
comparing to other flags every $f_i$ takes its minimal value (!). For
example, it was found that the minimal flag for $G_2$ models (both
rational and trigonometric) is ${\cal P}^{(1,2)}$ and the
characteristic vector (\ref{char.vec.}) is $(1,2)$
\cite{Rosenbaum:1998}.

Our final goal will be a search for minimal flags. It is worth to
mention that one of the situations, when several flags of invariant
subspaces can exist, occurs for the operator written in different
variables while invariant subspaces remain to be polynomial one.
Minimal flags have a remarkable property -- they are preserved by
corresponding trigonometric Hamiltonians if the latter are written in
appropriate variables.

A general strategy of our study is the following: (i) as a first
step we gauge rotate the ground state eigenfunction, (ii) choosing
a certain orbit we construct a particular set of variables which
lead to an algebraic form of the gauge-rotated Hamiltonian, (iii)
exploiting ambiguity in definition of the invariants of the fixed
degrees we search for variables preserving a minimal flag.

\setcounter{equation}{0}
\section{The rational $G_2$ model }

The rational $G_2$ model was introduced for the first time by
Wolfes \cite{Wolfes:1974} and later on was obtained in the
Hamiltonian Reduction method
\cite{Olshanetsky:1977,Olshanetsky:1983}. This model allows an
interpretation as a model of a three-identical particles with two-
and three-body interactions. It was extensively studied in
\cite{Rosenbaum:1998}, \cite{Turbiner:1999}.

The root system of the $G_2$ algebra is defined in 3-dimensional
space with a constraint to the hyperplane $y_1+y_2+y_3=0$. Six
positive roots are
\[
\{ e_2-e_1\ ,\ e_3-e_2\ ,\ -e_1+e_3\ ,\ e_1-2e_2+e_3\ ,\ -2e_1+e_2+e_3\ ,\
-e_1-e_2+2e_3 \}\ .
\]

The Hamiltonian of the rational $G_2$ model can be written in the
form
\begin{align}
 \label{e2.1}
 {\cal H}_{\rm G_2}^{(r)} &=
 \frac{1}{2}\sum_{k=1}^{3}\big[-\frac{\pa^{2}}{\pa x_{k}^{2}}
 + \om^2 x_k^2\big] + V_{G_2}(x) \ , \non\\
 V_{G_2}(x) &= g_s\sum_{k<l}^{3}\frac{1}{(x_{k} - x_{l})^2}
 + 3 g_l\sum_{ k<l \ ,\ k,l \neq m}^{3}
 \frac{1}{(x_{k} + x_{l}-2 x_{m})^2}  \ ,
\end{align}
where $\om$ is a frequency parameter and $g_s=\nu(\nu-1) > -
\frac{1}{4}$, $g_l=\mu (\mu -1) > - \frac{1}{4}$ are coupling
constants associated with the two-body and three-body interactions.
The existence of two coupling constants reflects the fact that the
root system contains two sets of roots, long and short: $\R_{short}$
with roots of length 2 and $\R_{long}$ with roots of length 6.
Parameter $\om$ in (\ref{e2.1}) is the only dimensional parameter in
the Hamiltonian and the eigenvalues should be proportional to it. A
connection to the root system with the model (\ref{e2.1}) appears when
the center-of-mass coordinate is separated out and the relative motion
is studied. Let us introduce the Perelomov coordinates of the relative
motion \cite{Perelomov:1971}
\begin{equation}
\label{e2.4}
 y_{1,2,3}^{Perelomov}=x_{1,2,3}-\frac{1}{3}X\ ,\ X= x_1 + x_2 + x_3\ ,
\end{equation}
where $x_i$ are the Cartesian coordinates of particles and $X$ is
the center-of-mass coordinate. Another convenient set of relative
variables is given by the standard Jacobi coordinates
\begin{equation}
\label{e2.4a}
 y_{1}^{Jacobi} = x_{2}-x_{1}\ ,\
 y_{2}^{Jacobi} = x_{3}-x_{2}\ ,\
 y_{3}^{Jacobi} = x_{1}-x_{3}\ .\
\end{equation}
Both sets of the relative coordinates obey a condition
$y_1+y_2+y_3=0$. Thus the relative motion can be studied in
three-dimensional $y$-space with the constraint $y_1+y_2+y_3=0$.
Transition from (\ref{e2.4}) to (\ref{e2.4a}) corresponds to the
interchange of two sets of roots -- ${\cal R}_{short}$ and ${\cal
  R}_{long}$.  One can identify the variables $y$ either with
$y^{Perelomov}$ or $y^{Jacobi}$. Due to the symmetry of the
Hamiltonian with respect to the interchange of long and short roots a
duality between the two sets of variables, $y^{Perelomov}$ and
$y^{Jacobi}$, appears. It will be demonstrated below.

The ground state of relative motion is given by
\begin{align}
\label{e2.2}
 \Psi_{0}^{({\rm r})}(y) &= (\De_1^{(r)}(y))^{\nu}
 (\De_2^{(r)}(y))^{\mu}
 \exp\{-\frac{\om}{2} \sum_{k=1}^3 y_k^2\} ~, \qquad y_1+y_2+y_3=0 ~, \non\\
 \qquad E_0\ =\ & \frac{3}{2}\om(1+2\nu+2\mu)\ ,
\end{align}
where $\De_1^{(r)}(y)$ and $\De_2^{(r)}(y)$ are Vandermonde determinants
\begin{align*}
\De_1^{(r)}(x) &=\prod_{\R_{short}} (\alpha_k\cdot y) 
= \prod_{j<i}^3 (y_i-y_j) \ , \non\\
\De_2^{(r)}(x) &=\prod_{\R_{long}} (\alpha_k\cdot y)
=\prod^3_{i<j; \ i,j\neq k}(y_i+y_j-2y_k)\ .
\end{align*}

Let us make a gauge rotation of the Hamiltonian (\ref{e2.1}) with
the ground state eigenfunction (\ref{e2.2}):
\begin{align}
\label{e2.3}
 h^{\rm (r)}_{G_2} = -2 (\Psi_0^{(r)}(y))^{-1}
 ({\cal H}^{(r)}_{G_2}-E_0) \Psi_0^{(r)}(y) \  ,
\end{align}
and separate out the center-of-mass motion.

Following the symmetries of the Hamiltonian (\ref{e2.1}) let us define
the $G_2$ Weyl-invariant polynomials by averaging over the simplest
orbit $\Om(e_2-e_1)$, generated by $(e_2-e_1)$. It has six elements:
\begin{equation}
\label{e_weyl}
 t^{(\Om)}_a(y) = \frac{1}{6} \sum_{k=1}^{6} (\om_k \cdot y)^a, \qquad
 \om_k \in \Om(e_2-e_1)\ ,
\end{equation}
(cf. (\ref{orbit})), where $a = 2,6$ are the degrees of the
$G_2$-algebra and $\om_k, k=1,2,\ldots 6$ are the orbit elements.
Explicitly,
\[
 t^{(\Om)}_2(y) \ = \ 2 (y_1^2+y_2^2+y_1y_2) \ ,
 \]
 \[
 t^{(\Om)}_6(y) \ = \
 105y_1^4y_2^2 + 66y_1^5y_2+66y_1y_2^5 + 22y_1^6 + 22y_2^6
 + 100y_1^3y_2^3 + 105y_1^2y_2^4\ ,
\]

Taking different orbits in the formula of averaging (\ref{e_weyl}) we
obtain different invariants. They are related to each other as
\begin{align}
\label{e2.8_orbit}
 t_2^{(\Om')} & =  t_2^{(\Om)} \ , \non \\
 t_6^{(\Om')} & =  t_6^{(\Om)} + A^{(6,\Om)} (t_2^{(\Om)})^3 \ .
\end{align}
up to multiplicative factors. The general invariants of the lowest
degree of the $G_2$-algebra can be found through invariants obtained
by averaging, they are algebraically related to $t_a^{(\Om)}$. In
fact, an arbitrary set of invariants generating the $S^{W_g}$ algebra
can be found through a particular orbit invariants and has a form:
\begin{align}
\label{e2.8t}
 t_2 & =  t_2^{(\Om)}(y) \ , \non \\
 t_6 & = t_6^{(\Om)}(y) + A^{(6)} (t_2^{(\Om)}(y))^3 \ .
\end{align}
It can be shown that by taking $t_{2,6}$ as new variables in
(\ref{e2.3})  we always arrive at the algebraic form of the
gauge-rotated rational $G_2$ Hamiltonian for any value of the
parameter $A^{(6)}$. One may assume that such a property should
correspond to the flag with characteristic vector $(1,3)$, which
is invariant with respect to transformation (\ref{e2.8t}). The
existence of this flag can be easily confirmed.

A set of flags preserved by the Hamiltonian in coordinates
(\ref{e2.8t}) can be obtained by analyzing its action on monomials
$\phi_{\vec{p}}=t_2^{p_1}t_6^{p_2}$ labelled by vectors
$\vec{p}=(p_1,p_2)$. The monomial $\phi_{\vec{p}}$ is mapped into
a sum of monomials $\phi_{\vec{p}-\vec{d}_i}$. There are among
$d_i$ two zero vectors $(0,0)$ corresponding to terms $-4\om t_1
\pa_{t_1}$ and $-12\om t_2\pa_{t_2}$, which determine eigenvalues
of the Hamiltonian. Other shift vectors are $(-1,0)$, $(2,-1)$ and
$(5,-2)$. It means that the minimal flag for the general coordinates
(\ref{e2.8t}) is $(2,5)$.

However, one can tune the value of the parameter $A^{(6)}$ to
eliminate the term $(5,-2)$ in the Hamiltonian. Remarkably there exist
two such values of this parameter,
\begin{align}
\label{e2.8}
 A^{(6)} = -9/4 \quad \text{and} \quad A^{(6)} = -11/4 \ ,
\end{align}
resulting in smaller flag $(1,2)$ (see below the equation
(\ref{e2.6f})). Following the definition of Section 2 this flag
${\cal P}^{(1,2)}$ is {\it minimal}.~\footnote{It is
remarkable that this flag is invariant under the action of the 
trigonometric $G_2$ Hamiltonian, when this Hamiltonian is written
in appropriate variables.}

Making the substitution
\begin{align}
\label{var1}
 \ta_2 & =  t_2^{(\Om)}(y) \ , \non \\
 \ta_6 & = t_6^{(\Om)}(y) -\frac{9}{4}(t_2^{(\Om)}(y))^3 \ .
\end{align}
we arrive at the Hamiltonian
\begin{align}
 \label{e2.5}
 h_{\rm G_2}^{(r,1)} & = 4\ta_2\pa^2_{\ta_2\ta_2}
 +24 \ta_6\pa^2_{\ta_2\ta_6}
 +18\ta_2^2\ta_6\pa^2_{\ta_6\ta_6} \non\\[2mm]
 &~~ -\left\{4\om\ta_2-4[1+3(\mu+\nu)]\right\}\pa_{\ta_2}
 -\left[ 12\om\ta_6-9(1+2\nu)\ta_2^2\right]\pa_{\ta_6}\ .
\end{align}
This is an {\it algebraic} form of the rational $G_2$ model. It
can be easily checked that the Hamiltonian (\ref{e2.5}) has
infinitely-many finite-dimensional invariant subspaces
\begin{equation}
\label{e2.6}
 {\cal P}_n^{(1,2)} \ = \ \langle \ta_2^{p_1} \ta_6^{p_2}
 |\ 0 \leq p_1 + 2p_2 \leq n \rangle\ ,\ n=0,1,\ldots\ ,
\end{equation}
with the characteristic vector
\begin{equation}
\label{e2.6f}
 \vec f \ =\ (1,2)\ ,
\end{equation}
forming the minimal flag ${\cal P}^{(1,2)}$ of the rational $G_2$
model which we denote ${\cal P}^{(G_2)}={\cal P}^{(1,2)}$ (see a
discussion above).

The Hamiltonian (\ref{e2.5}) admits a representation in terms of a
non-linear combination of the generators of some
infinite-dimensional finitely-generated algebra $g^{(2)}$ (for
definition and discussion see \cite{Rosenbaum:1998}). It leads to
the $g^{(2)}$-{\it Lie-algebraic} form of the rational $G_2$
model. This form depends on a subset of the generators of the
$gl_2 \ltimes R^3$-subalgebra of algebra $g^{(2)}$ only and does
not contain a raising generator of $g^{(2)}$. The generators of
$g^{(2)}$ with excluded raising generator have infinitely-many
common finite-dimensional spaces which coincide to the
finite-dimensional invariant subspaces of the Hamiltonian
(\ref{e2.5}).

We already studied the Hamiltonian (\ref{e2.3}) corresponding to
the first choice of variables, when $A^{(6)} = -9/4$ in
(\ref{e2.8t}). Now we explore the second choice, when $A^{(6)} =
-11/4$. One can easily represent these variables in terms of the
variables of the first choice
\begin{align}
\label{var2}
 \tta_2(y) & = t^{(\Om)}_2(y) = \ta_2 \ , \non \\
 \tta_6(y) &= -t^{(\Om)}_6(y) + \frac{11}{4} (t^{(\Om)}_2(y))^3
 = - \ta_6 +\frac{1}{2} \ta_2^3\ ,
\end{align}
where $y$'s are the Perelomov coordinates (\ref{e2.4}). Making the
change of variables (\ref{var2}) in (\ref{e2.5})
the Hamiltonian (\ref{e2.3}) takes the form
\begin{align}
\label{e2.11}
 h_{\rm G_2}^{(r,2)} & = 4\tta_2\pa^2_{\tta_2\tta_2}
        +24 \tta_6\pa^2_{\tta_2\tta_6}
        +18\tta_2^2\tta_6\pa^2_{\tta_6\tta_6} \non\\[2mm]
 &~~ -\left\{4\om\tta_2-4[1+3(\mu+\nu)]\right\}\pa_{\tta_2}
 -\left[ 12 \om\tta_6-9(1+2\mu)\tta_2^2\right]\pa_{\tta_6}\ .
\end{align}
(cf. (\ref{e2.5})). This is another {\it algebraic} form of the
rational $G_2$ model, which is dual to (\ref{e2.5}): the
Hamiltonian (\ref{e2.11}) differs from (\ref{e2.5}) by permutation
$\nu\leftrightarrow\mu$ only. It reflects a symmetry between sets
of short and long roots. It is evident that the Hamiltonian
(\ref{e2.11}) preserves the flag ${\cal P}^{(1,2)}$ and admits a
representation in terms of the generators of the algebra $g^{(2)}$
but now being written in the coordinates $\tta$'s where
$\nu\leftrightarrow\mu$.

The duality between short and long roots also occurs when we make
the identification of $y$ variables either with the Perelomov
coordinates (\ref{e2.4}) or with the Jacobi coordinates
(\ref{e2.4a}). It appears in a relation between the
$\ta$-variables (of the first choice of $A^{(6)}$, see
(\ref{e2.8})) in the Perelomov coordinates and the
$\tta$-variables (of the second choice of $A^{(6)}$, see
(\ref{e2.8})) in the Jacobi coordinates and visa versa, e.g.
\begin{align}
\label{Per_Jac}
 \ta_2(y^{Jacobi}) =  3 \tta_2(y^{Perelomov})~, \qquad
 \ta_6(y^{Jacobi}) = 27 \tta_6(y^{Perelomov})~.
\end{align}

It is worth emphasizing that the coordinates $\tta$'s (\ref{var2})
have a remarkable property: they are unique coordinates which can be
`trigonometrized' -- they coincide with the rational limit of certain
trigonometric coordinates, $\lim_{\al \rar 0} \dfrac{\sin \al
  x}{\al}=x$, in which the trigonometric $G_2$ model gets an algebraic
form (see \cite{Rosenbaum:1998}). The $\ta_{2,\,6}$-coordinates do not
have such a property \footnote{ It is worth noting that in the
  coordinates $(\ta_2, \sqrt{\ta_6})$ the gauge-rotated rational $G_2$
  Hamiltonian (\ref{e2.3}) can be rewritten in terms of the generators
  of the maximal affine subalgebra of the $gl(3)$-algebra
  \cite{Rosenbaum:1998}. Thus, this Hamiltonian possesses two
  different hidden algebras: the $g^{(2)}$ algebra acting on the
  configuration space parameterized by the $\ta(\tta)$-coordinates and
  the $gl(3)$-algebra acting on the configuration space parameterized
  by the $(\ta_2, \sqrt{\ta_6})$ coordinates (see
  \cite{Rosenbaum:1998} and Eqs.(2.6)-(2.7) therein).}.

It is interesting to analyze a transformation preserving the flag
${\cal P}^{(G_2)}$. The most general polynomial transformation
which preserve the linear space ${\cal P}_{n}^{(1,2)}$
(\ref{e2.6}) is
\begin{align}\label{e2.15t}
 \ta_2 &\rar \ta_2 \ , \non\\
 \ta_6 &\rar \ta_6 \ +\ a^{(2)} \ta_2^2 \ ,
\end{align}
where $a^{(2)}$ is an arbitrary number of the dimension $[\om^{-1}]$
for any $n$. It is evident that this transformation preserves the
whole flag ${\cal P}^{(G_2)}$. We find that there exist two algebraic
operators which preserve the same flag ${\cal P}^{(G_2)}$. However,
the variables $\ta$'s and $\tta$'s in which these operators have being
written are related one to each other through a transformation of the
type (\ref{e2.15t}) (see the relation (\ref{var2})). Hence these two
algebraic operators are non-equivalent. One suspects that the fact of
the existence of two non-equivalent algebraic forms of the rational
$G_2$ model (which look very much alike) reflects a certain intrinsic
degeneracy of the model and should not hold for the trigonometric
case. A study of the trigonometric $G_2$ model confirms that there
exists a single $g^{(2)}$-Lie-algebraic form.

Thus, the operators (\ref{e2.5}), (\ref{e2.11}) possess infinitely
many finite-dimensional invariant subspaces. These invariant subspaces
coincide with the finite-dimensional representation spaces of the
algebra $g^{(2)}$. It is worth noting that for $\om=\nu=\mu=0$ both
operators $h^{(r,1(2))}_{G_2}$ coincide and represent the flat space
Laplacian written in the $g^{(2)}$-Lie-algebraic form. It is evident
that in this case there are no polynomial eigenfunctions in $\ta
(\tta)-$coordinates.

Both operators (\ref{e2.5}), (\ref{e2.11}) are triangular in the basis
of monomials $\ta_2^{p_1}\ta_6^{p_2}$ ($\tta_2^{p_1}\tta_6^{p_2}$).
Therefore, the spectrum of (\ref{e2.5}), (\ref{e2.11}), $h^{\rm
  (r)}_{G_2} \varphi = - 2\ep \varphi$, can be found explicitly and is
equal to
\begin{align}
\label{e2.15}
  \ep_{n_1,n_2}\ =\ \om (2 n_1 + 6 n_2) \ ,
\end{align}
where $n_i$ are non-negative integers $n_1, n_2=0,1,\ldots$
(coefficients 2 and 6 are degrees of $G_2$). The spectrum does not
depend on the coupling constants $g_l$, $g_s$ (but the reference
point for energy (\ref{1.3}) does), it is equidistant and
corresponds to the spectrum of two harmonic oscillators with
frequencies $2\om$ and $6\om$. Degeneracy of the spectrum is
related to the number of partitions of integer number $n$ to two
weighted integers $n_1 + 3 n_2$. The spectrum of the original
rational $G_2$ Hamiltonian (\ref{e2.1}) is $E_n=E_0+\ep_n$. It is
worth noting that the Hamiltonian (\ref{e2.11}) (as well as
(\ref{e2.5})) possesses a remarkable property: there exists a
family of eigenfunctions which depend on the single variable
$\tta_2 (\ta_2)$. These eigenfunctions are the associated Laguerre
polynomials. This property allows to construct a
quasi-exactly-solvable generalization of the rational $G_2$ model.
It will be done elsewhere.

It is worth mentioning that the boundaries of configuration space are
determined by zeros of the ground state wave function (\ref{e2.2}). In
$\ta$-variables it is the algebraic curve
\begin{align}
\label{e2.16a}
 \left( \De_1 \De_2 \ (\ta) \right)^2 =
 - 27 \ta_6(\ta_6 - 1/2 \ta_2^3)\ =\ 0\ ,
\end{align}
and in $\tta$-variables it has the same form
\begin{align}
\label{e2.16b}
 \left( \De_1 \De_2 \ (\tta) \right)^2 =
 - 27 \tta_6(\tta_6 - 1/2 \tta_2^3)\ =\ 0\ .
\end{align}
In $\ta_6, \tta_6$ variables associated with the Perelomov
$y$-coordinates the product of discriminants looks especially
simple
\begin{align}
\label{e2.16c}
  \left( \De_1 \De_2 \ (y) \right)^2 = 27 \ta_6 (y)  \tta_6 (y) ~,
\end{align}
where symmetry between short and long roots is seen explicitly. Also
note \cite{Bourbaki}
\begin{align}
\label{e2.jac}
 \left[\det\left( \frac{\pa  \ta_i}{\pa y_k} \right)\right]^2 =
 16\ \left( \De_1 \De_2 \right)^2  \ .
\end{align}


\setcounter{equation}{0}
\section{The rational $F_4$ model}

The Hamiltonian of the rational $F_4$ model written in the basis of
the standard $F_4$ roots has the form (see \cite{blt} \footnote{There
  is a misprint in \cite{blt} in the definition of coupling constants
  for the Hamiltonian (3.1): it should read $g_1=\mu(\mu-1)>-1/4 ~, ~
  g=(1/2)\nu(\nu-1)>-1/8 $~. }),
\begin{align}
\label{e3.1}
 {\cal H}_{\rm F_4}^{(r)} = &\ \frac{1}{2}\
\sum_{i=1}^{4} \left( -\pa_{x_i}^2 + \om^2 x_i^2 \right) +
g_l \sum_{j>i}^4 \left( \frac{1}{(x_i-x_j)^2} + \frac{1}{(x_i+x_j)^2}\right)\\
&+ \frac{g_s}{2} \sum_{i=1}^{4}\frac{1}{{x_i}^2} + 2g_s
\sum_{ \nu's=0,1} \frac{1}{ \left[ x_1 + (-1)^{\nu_2}x_2+
(-1)^{\nu_3}x_3+ (-1)^{\nu_4}x_4 \right]^2}\non\ ,
\end{align}
where $g_l=\nu(\nu-1)$, $g_s =\mu(\mu-1)$ are coupling constants
related to sets of long and short roots, ${\cal R}_{long}$ and
${\cal R}_{short}$, correspondingly, and $\om$ is a frequency. Its ground
state can be written as
\begin{align}
\label{e3.2}
 & \Psi_{0}^{({\rm r})}(x) = \left(\De_-\De_+\right)^{\nu}
 \left(\De_0 \De \right)^{\mu}
\exp\left( - \frac{\om}{2} \sum_{i=1}^{4} {x_i}^2\right)\ ,
\end{align}
where
\begin{align}
\label{e3.3}
 \De_{+}\De_{-} &= \prod_{\R_{long}} (\alpha_k\cdot x)  =
 \prod_{j<i}^4 (x_i+ x_j) \prod_{j<i}^4 (x_i- x_j) \ , \non\\
 \De_{0}\De~ &= \prod_{\R_{short}} (\alpha_k\cdot x)  \non\\
 &= \prod_{i=1}^4 x_i \prod_{\nu's=0,1}
 \left(\frac{x_1 + (-1)^{-\nu_2}x_2 +
 (-1)^{-\nu_3}x_3+ (-1)^{-\nu_4}x_4 }{2}\right)\ .
\end{align}
The ground state energy is
\begin{equation}
\label{e3.4}
 E_0= 2\om (1 + 6\mu+ 6\nu) \ .
\end{equation}

Let us make gauge rotation of the Hamiltonian (\ref{e3.1}) with
the ground state eigenfunction (\ref{e3.2}) as a gauge factor
\begin{equation}
\label{e3.6}
 h_{\rm F_4}^{(r)} = -2 (\Psi_0^{(r)}(x))^{-1}
 ({\cal H}_{\rm F_4}^{(r)}-E_0) \Psi_0^{(r)}(x)\ .
\end{equation}
As new variables we take Weyl invariant polynomials of the {\it
lowest} degrees of the group $W_{F_4}$ found by averaging
elementary polynomials $(\om\cdot x)^{a}$ over the 24-element orbit
$\Om$ generated by the root $(e_1 + e_2)$
\begin{align}
\label{orbit_f4}
 t_{a}^{(\Om)}(x) = \frac{1}{12}\sum_{\om\in\Om} (\om \cdot x)^{a}\ ,
 \qquad a=2,6,8,12 \ .
\end{align}
(cf. (\ref{orbit})). In this case the powers $a=2,6,8,12$ are the
{\em degrees} of the group $W_{F_4}$. As we mentioned in Section
2 the invariants of the fixed degrees $a$  are defined
ambiguously, up to some non-linear combinations of invariants of
the lower degrees
\begin{align}
\label{e3.9t}
 t_2^{(\Om)} &\rar  t_2^{(\Om)} \ , \non \\
 t_6^{(\Om)} &\rar  t_6^{(\Om)} + A^{(6)} (t_2^{(\Om)})^3 \ , \non\\
 t_8^{(\Om)} &\rar  t_8^{(\Om)} + A^{(8)}_1 (t_2^{(\Om)})^4 + A^{(8)}_2 t_2^{(\Om)}
 t_6^{(\Om)} \ ,\non\\
 t_{12}^{(\Om)} &\rar  t_{12}^{(\Om)} + A^{(12)}_1 (t_2^{(\Om)})^6+
 A^{(12)}_2 (t_2^{(\Om)})^3 t_6^{(\Om)} +
 A^{(12)}_3 (t_2^{(\Om)})^2 t_8^{(\Om)} + A^{(12)}_4 (t_6^{(\Om)})^2 \ ,
\end{align}
where $A^{(6,8,12)}_i$ are arbitrary parameters. It can be shown
that operator $h_{\rm F_4}^{(r)}$ is algebraic and it preserves a
flag of polynomials for arbitrary values of these parameters.
There are two flags $(2,6,8,12)$ and $(2,6,8,11)$ which are
invariant under transformations (\ref{e3.9t}). They are similar to
the flags $(1,3)$ and $(2,5)$ for the $G_2$ case.

Then we must search for a minimal flag. From a technical point of view
it can be found by eliminating certain terms in the Hamiltonian, by
choosing appropriate parameters $A$'s. In \cite{blt} the minimal flag,
denoted as ${\cal P}^{(F_4)}$, is found. This flag is generated
by the  spaces of quasi-homogeneous polynomials
\begin{equation}
\label{e3.10}
 {\cal P}_{n}^{(1,2,2,3)} =
\langle  \ta_1^{p_1} \ta_3^{p_3} \ta_4^{p_4} \ta_6^{p_6} | \ 0
\leq p_1 + 2 p_3 + 2 p_4 + 3 p_6 \leq n \rangle\ ,
\end{equation}
with the characteristic vector $\vec f\ $
\begin{equation}
\label{e3.11f}
 \vec f = (1,2,2,3)\ ,
\end{equation}
which give rise to the flag ${\cal P}^{(1,2,2,3)}$. Hence, ${\cal
P}^{(F_4)} = {\cal P}^{(1,2,2,3)}$ is the minimal flag for the
rational $F_4$ model.

The characteristic vector (\ref{e3.11f}) coincides with the highest
root among short roots in the $F_4$ root system written in the basis
of simple roots \footnote{We thank Victor Ka\^c for this remark. The
  similar statement holds for $A_n$-Calogero-Sutherland models where
  the flags are characterized by $\vec f \ =\ (1,1,\ldots, 1)$
  \cite{Ruhl:1995} and rational-trigonometric $G_2$ models where $\vec
  f \ =\ (1,2)$ \cite{Rosenbaum:1998}. However, it is not the case for
  the rational $E_{8}$ model (see below).}.  Note that the most
general polynomial transformation preserving the linear space ${\cal
  P}_{n}^{(1,2,2,3)}$ is
\begin{align}\label{e3.11t}
 s_1 &\rar s_1 \ , \non\\
 s_2 &\rar s_2\ +\ a_2 s_1^2 + b_2 s_3\ , \non\\
 s_3 &\rar s_3\ +\ a_3 s_1^2 + b_3 s_2\ , \non\\
 s_4 &\rar s_4\ +\ a_4 s_1^3 + b_4 s_1 s_2 + c_4 s_1 s_3\ ,
\end{align}
where $a,b,c$ are arbitrary numbers.

Explicitly the  set of variables preserving the minimal flag
${\cal P}^{(1,2,2,3)}$ found in Ref. \cite{blt} is:
\begin{align}
\label{e3.12var1}
 \ta_2 &= t_2^{(\Om)} \ ,\non\\
 \ta_6 &= \frac{1}{12} t_6^{(\Om)}-\frac{1}{12}\big(t_2^{(\Om)}\big)^3 \ ,
\non\\
 \ta_8 &= \frac{1}{80} t_8^{(\Om)} -\frac{1}{30} t_2^{(\Om)} t_6^{(\Om)}
  + \frac{1}{48} \big(t_2^{(\Om)}\big)^4 \ ,\non\\
 \ta_{12} &=
 \frac{1}{720} t_{12}^{(\Om)} - \frac{5}{288} \big(t_2^{(\Om)}\big)^2
 t_8^{(\Om)} + \frac{1}{27} \big(t_2^{(\Om)}\big)^3 t_6^{(\Om)} -
 \frac{29}{1440} \big(t_2^{(\Om)}\big)^6 - \frac{1}{1080} \big(t_6^{(\Om)}\big)^2\ ,
\end{align}
In these variables the algebraic form of the gauge-rotated
Hamiltonian (\ref{e3.6}) is:
\begin{align}
\label{e3.14}
 h_{\rm F_4}^{(r,1)} = &
4 \ta_2 \frac{\pa^2}{\pa \ta_2^2} + \frac{2}{3} (10\ta_2 \ta_8 +
{\ta_2}^2 \ta_6) \frac{\pa^2}{\pa \ta_6^2} + 2(\ta_2 \ta_{12} +
2\ta_8 \ta_6)\frac{\pa^2}{\pa \ta_8^2}\non\\
&\mbox{} +24\ta_6 \frac{\pa}{\pa \ta_2} \frac{\pa}{\pa \ta_6}
+32\ta_8 \frac{\pa}{\pa \ta_2} \frac{\pa}{\pa \ta_8}
+48\ta_{12}\frac{\pa}{\pa \ta_2} \frac{\pa}{\pa \ta_{12}}\non\\
&\mbox{} + \frac{8}{3} ({\ta_2}^2 \ta_8 + 6\ta_{12}) \frac{\pa}{\pa
\ta_6}\frac{\pa}{\pa \ta_8} + 4({\ta_2}^2 \ta_{12} + 8{\ta_8}^2)
\frac{\pa}{\pa \ta_6} \frac{\pa}{\pa \ta_{12}}\non\\
&+ 4 (3 \ta_6 \ta_{12} + 2 \ta_2 {\ta_8}^2) \frac{\pa}{\pa \ta_8}
\frac{\pa}{\pa \ta_{12}} + 6(2{\ta_8}^2 \ta_6 + \ta_2 \ta_8
\ta_{12}) \frac{\pa^2}{\pa \ta_{12}^2} \non\\
&\mbox{} - 4[\om \ta_2 - 2(6\nu +6 \mu +1)]\frac{\pa}{\pa \ta_2}
-[12 \om \ta_6 - {\ta_2}^2 (4\nu +2 \mu + 1)]\frac{\pa}{\pa \ta_6}\non\\
&- 4[4\om {\ta_8}-{\ta_6}(1+3\nu)]\frac{\pa}{\pa \ta_8}-4[6\om
\ta_{12}-\ta_2 \ta_8 (2+3\nu)]\frac{\pa}{\pa \ta_{12}}
\end{align}
The variables (\ref{e3.12var1}) are remarkable as they are the
rational limits of certain trigonometric variables in which
the trigonometric $F_4$ model takes an algebraic form \cite{blt}.

However, the set of variables (\ref{e3.12var1}) does not exhaust all
the possible sets of variables leading to the minimal flag
(\ref{e3.11f}). Similarly to what happens with the rational $G_2$
model, there exists one more set of variables
\begin{align}
\label{e3.15var2}
 {\tta_2} &= t_2^{(\Om)} =\ta_2 \ ,\non\\[2mm]
 {\tta_6} &= - \frac{1}{12}t_6^{(\Om)} + \frac{1}{8}\big(t_2^{(\Om)}\big)^{3}
 = -\ta_6 +\frac{1}{24}\ta_2^3 \ ,\non\\[2mm]
 {\tta_8} &= \frac{1}{80}t_8^{(\Om)} - \frac{13}{240} t_2^{(\Om)}t_6^{(\Om)}
 +\frac{3}{64} \big(t_2^{(\Om)}\big)^4
  = \ta_8-\frac{1}{4}\ta_2\ta_6+\frac{1}{192}\ta_2^4\ ,\non\\[2mm]
 {\tta_{12}} &=  -\frac{1}{720} t_{12}^{(\Om)} +
 \frac{61}{17280}\big(t_6^{(\Om)}\big)^2
 + \frac {109}{5760} \big(t_2^{(\Om)}\big)^2 t_8^{(\Om)}
 -\frac{847}{17280} \big(t_2^{(\Om)}\big)^3 t_6^{(\Om)} +
 \frac{109}{3840}\big(t_2^{(\Om)}\big)^6
 ~~~\non\\
 &= -\ta_{12}+\frac{1}{8}\ta_2^2\ta_8+\frac{3}{8}\ta_6^2 -
 \frac{1}{32}\ta_6\ta_2^3+\frac{1}{2304}\ta_2^6 ~,
\end{align}
(cf. (\ref{e3.12var1})) leading to an algebraic form of $h_{\rm
F_4}^{(r)}$ which preserves the minimal flag ${\cal P}^{(F_4)}$.
The explicit expression for the Hamiltonian in the variables
(\ref{e3.15var2}) gets the same form as (\ref{e3.14})  with
 $\nu$ and $\mu$ exchanged:
\begin{align}
\label{e3.16}
 h_{\rm F_4}^{(r,2)}(\tta) =
 h_{\rm F_4}^{(r,1)}(\ta; \nu\leftrightarrow \mu)
\end{align}
Clearly the Hamiltonian continues to be algebraic under
the transformations (\ref{e3.11t}). 

We were able to find the one-parametric algebra of differential
operators for which ${\cal P}^{(F_4)}_n$ is the space of
finite-dimensional irreducible representation (see Appendix B in
\cite{blt}).  Furthermore, the finite-dimensional representation
spaces arise for different integer values of the parameter. They form
an infinite non-classical flag which coincides with ${\cal P}^{(F_4)}$
(\ref{e3.10}). We call this algebra $f^{(4)}$. Like the algebra
$g^{(2)}$ introduced in \cite{Rosenbaum:1998} in relation to the
$G_2$ models, the algebra $f^{(4)}$ is infinite-dimensional yet
finitely-generated. The rational $F_4$ Hamiltonian in either algebraic
form (\ref{e3.14}) or (\ref{e3.16}) can be rewritten in terms of the
generators of this algebra.

The variables (\ref{e3.12var1}) and (\ref{e3.15var2}) can not be
related by the transformation (\ref{e3.11t}). Therefore these two
$f^{(4)}$-Lie-algebraic forms are non-equivalent from the point of
view of the transformation (\ref{e3.11t}). We interpret this fact by a
certain intrinsic degeneracy of the rational $F_4$ model. When the
trigonometric $F_4$ model is considered one can show that there exists
the only one $f^{(4)}$-Lie-algebraic form (up to a transformation
(\ref{e3.11t})) \cite{blt}.

Thus, the operators (\ref{e3.14}),(\ref{e3.16}) have infinitely many
finite-dimensional invariant subspaces. The finite-dimensional
representations of the algebra $f^{(4)}$. If $\om=\nu=\mu=0$, the
operators $h^{(r,1(2))}_{F_4}$ coincide and become the flat space
Laplacian written in the $g^{(2)}$-Lie-algebraic form, with no
polynomial eigenfunctions in $\ta (\tta)-$coordinates.

The operator (\ref{e3.14}) (as well as (\ref{e3.16})) is 
triangular in the basis of monomials
$\ta_2^{p_1}\ta_6^{p_2}\ta_8^{p_3}\ta_{12}^{p_4}$. One can find
the spectrum of (\ref{e3.14}), $h_{\rm F_4}^{(r,1)} \varphi =
-2\ep\varphi$, explicitly
\begin{equation}
\label{e3.17}
  \ep_{n_1,n_2,n_3,n_4}=  \om (2 n_1 + 6 n_2 + 8 n_3+ 12 n_4) \ ,
\end{equation}
where $n_i=0,1,\ldots$ are non-negative integers. Degeneracy of the
spectrum is equal to the number of partitions of an integer number $n$
to four weighted integers $n_1+3n_2+4n_3+6n_4$. The spectrum does not
depend on the coupling constants $g_l$, $g_s$, it is equidistant and
coincides (with different degeneracy) with the spectrum of the
harmonic oscillator as well as with that of the rational $D_4$ model.
Finally, the energies of the original rational $F_4$ Hamiltonian
(\ref{e3.1}) are $E_n=E_0+\ep_n$. It is worth noting that the
Hamiltonian (\ref{e3.14}) possesses a remarkable property: there
exists a family of eigenfunctions which depend on the single variable
$\ta_2$. These eigenfunctions are the associated Laguerre polynomials.
As an illustration the first eigenfunctions are presented in the
Appendix A.  Again, this property allows to construct a
quasi-exactly-solvable generalization of the rational $F_4$ model. It
will be done elsewhere.

Configuration space of the rational $F_4$ model (\ref{e3.1}) is
defined by zeros of the ground state eigenfunction, i.e. by zeros of
the pre-exponential factor in (\ref{e3.2}). These zeros also define
boundaries of the Weyl chamber (see e.g.  \cite{Olshanetsky:1983}).
The pre-exponential factor (\ref{e3.2}) at $\nu=\mu=2$ can be written
as a product of two factors. The first is
\begin{equation}
\label{e3.18}
 \big(\Delta_+\,\Delta_- (\ta)\big)^{2}= - 192\, \ta_{12}^2 + 256\, \ta_8^3 \ ,
\end{equation}
it corresponds to the rational $D_4$ model which occurs for $g_s=0$ in
(\ref{e3.1}). The second one
\begin{align}
\label{e3.19}
 \big(\Delta_0\,\Delta (\ta)\big)^{2} = & \frac{1}{4096}
 (- 192\, \ta_{12}^2 + 256\, \ta_8^3 + 144\, \ta_6^2\, \ta_{12}
 - 27\, \ta_6^{4} - 192\, \ta_2\, \ta_6\, \ta_8^2 +
 48\, \ta_2^2\, \ta_8\, \ta_{12}
 \non \\[3pt] &
 + 30\, \ta_2^2\, \ta_6^2\, \ta_8
- 12 \ta_2^3\, \ta_6\, \ta_{12} + \frac{1}{2} \ta_2^3\, \ta_6^3 +
\ta_2^{4}\, \ta_8^2 - \frac{1}{2} \ta_2^5\, \ta_6\, \ta_8 +
\frac{1}{6} \ta_2^6\, \ta_{12}) \ ,
\end{align}
corresponds to the case of the degenerate $F_4$ model, $g_l=0$ (which
is equivalent to the $D_4$ model in dual variables, see
\cite{blt}). Thus, a boundary of the configuration space of the
rational $F_4$ model is the union of the algebraic hyper-surfaces
(\ref{e3.18})--(\ref{e3.19}) of degree 3 and 7 respectively,
being the reduced algebraic hyper-surface of degree 10.

In terms of the second set of variables one gets symmetric
expressions
\begin{eqnarray*}
(\Delta_+ \Delta_- (\tta))^2 &=&  - 192\,\tta_{12}^2 + 256\,\tta_8^3
+144\,\tta_6^2\,\tta_{12} - 27\,\tta_6^4 - 192\,\tta_2\,\tta_6
\,\tta_8^2 + 48\,\tta_2^2\, \tta_8\,\tta_{12}
\\  &&
+ 30\,\tta_2^2\,\tta_6^2\,\tta_8 - 12\,\tta_2^3\,\tta_6\,\tta_{12}
+ \frac{1}{2} \,\tta_2^3\,\tta_6^3 + \tta_2^4\,\tta_8^2 -
\frac{1}{2} \,\tta_2^5\,\tta_6\,\tta_8 + \frac{1}{6}
\,\tta_2^6\,\tta_{12}\ ,
\end{eqnarray*}
(cf.(\ref{e3.18})) and
\[
(\Delta\Delta_0 (\tta))^2 =  \frac{1}{4096}(- 192\,\tta_{12}^2 +
256\,\tta_8 ^3)\ ,
\]
(cf.(\ref{e3.19})).

In agreement with the  general theory (see e.g. \cite{Bourbaki}) 
\begin{align}
\label{e3.20jac}
 \left[\det\left( \frac{\pa \ta_a}{\pa x_k} \right)\right]^2
 = {4096} \big( \De_+ \De_- \big)^2 \big( \De_0 \De \big)^2  \ .
\end{align}

To conclude a discussion of the rational $F_4$ model, one can
state that the rational $F_4$ model admits two non-equivalent
algebraic and $f^{(4)}$-Lie-algebraic forms, correspondingly.
However, it does not manifest anything non-trivial. The existence
of these two set of variables is related to a phenomenon of
duality discussed in \cite{blt}: it can be easily checked that
taking the variables (\ref{e3.12var1}) and substituting in
(\ref{orbit_f4}) the variables $x$'s by $z$'s,
\begin{align}
\label{e3.21}
 z_{1,2}=x_1 \pm x_2 \quad ,\quad  z_{3,4}=x_3 \pm x_4
\end{align}
we get the variables (\ref{e3.15var2}).


\setcounter{equation}{0}
\section{The rational $E_6$ model}

The Hamiltonian of the rational $E_6$ model is built using the
root system of the $E_6$ algebra. A convenient way to represent
the Hamiltonian is to write it in an $8-$dimensional space
$\{x_1,x_2,\ldots x_8\}$ while imposing  two constraints
$x_7=x_6, \, x_8=-x_6$,
\begin{equation}
\label{e4.1}
  {\cal H}_{E_6} = -\frac{1}{2} \Delta^{(8)} +\frac{\om^2}{2}\ \sum_{i=1}^{8} x_i^2
  + V_{E_6} ~,
\end{equation}
where
\begin{align}
\label{e4.2}
 V_{E_6}(x) &= g \sum_{j<i =1}^{5} \left[
\frac{1}{(x_i +x_j)^2} + \frac{1}{(x_i - x_j)^2} \right]  \\
&+ g \sum_{\{\nu_j\}} \frac{1}{\left[\dfrac{1}{2}\left({ -x_8 +
x_7 +x_6 - \sum_{j=1}^5 (-1)^{\nu_j}x_j}\right)\right]^2} ~, \
&(\nu_j=0,1; \sum_{j=1}^{5} \nu_j \text{~is even} ) \non
\end{align}
is root-generated part of the potential with a coupling constant
$g=\nu(\nu-1)$. The configuration space is given by the principal
$E_6$ Weyl chamber.

In order to resolve the constraints, we introduce new variables
\begin{eqnarray}
\label{e4.3}
 y_i &=& x_i\ , \quad i=1\ldots 5 \non \\
 y_6 &=& x_6 + x_7 - x_8\ , \qquad  \mbox{
(using the constraint $y_6=3x_6$)} \non\\
 y_7 &=& x_6 -x_7\ ,\qquad \mbox{(using the constraint $y_7=0$)}
\non\\
 y_8 &=& x_6 +x_8\ ,\qquad \mbox{(using the constraint
$y_8=0$)}
\end{eqnarray}
in which the Laplacian becomes
\begin{equation}
\label{e4.4}
 \Delta^{(8)} = \Delta_y^{(5)} + 3 \frac{\pa^2}{\pa y_6^2} +
 2\left[
  \frac{\pa^2}{\pa y_7^2} + \frac{\pa^2}{\pa y_8^2} +
  \frac{\pa^2}{\pa y_7 \pa y_8} \ ,
  \right]
\end{equation}
while the potential part of (\ref{e4.1}) depends on $\{y_1 \ldots
y_6 \}$ only:
\begin{eqnarray}
\label{e4.5}
 V &=&  \frac{\om^2}{2} \left\{
 \sum_{i=1}^{5} y_i^2 + \frac{y_6^2}{3} \right\} +  g \, \sum_{j<i =1}^{5}
 \left[
\frac{1}{(y_i + y_j)^2} + \frac{1}{(y_i - y_j)^2}
 \right] \non \\
&& \, + \, g \sum_{\nu_j,j=1}^5
 \frac{1}{\left[\frac{1}{2}\left({ y_6 - \sum_{j=1}^{5}
(-1)^{\nu_j}y_j }\right)\right]^2}\ .
\end{eqnarray}
In this formalism imposing constraints implies a study of
eigenfunctions having no dependence on $y_7, y_8$. Hence,
$y_{7,8}$-dependent part of the Laplacian standing in square
brackets in (\ref{e4.4}) can be dropped off.

The ground state eigenfunction has a form
\begin{equation}
\label{e4.6}
 \Psi_0 = (\De_+^{(5)} \De_-^{(5)})^\nu \De_{E_6}^\nu
 {\rm e}^{-\frac{1}{2} \om \left\{
 \sum_{i=1}^{5} y_i^2 + \frac{y_6^2}{3} \right\}} \ ,
 \ E_0 = 3\om (1+12\nu )
\end{equation}
where
\begin{eqnarray*}
 \De_\pm^{(5)} &=& \prod_{j<i =1}^{5} (y_i \pm y_j)\\
 \De_{E_6} &=& \prod_{\{\nu_j\}}
 \left(y_6 + \sum_{j=1}^{5} (-1)^{\nu_j}y_j \right)
\end{eqnarray*}
with $g = \nu (\nu-1)$.

In order to find variables leading to the algebraic form of
gauge-rotated Hamiltonian,
\begin{equation}
\label{e4.6h}
 h_{E_6}^{(r)} (y_1\ldots y_6)\ =\ -2{\Psi_0}^{-1}
 ({\cal H}_{E_6}-E_0)(y_1\ldots y_6){\Psi_0}\ ,
\end{equation}
let us define a basis in the form of the Weyl-invariant polynomials
averaged over the 27-element orbit generated by the vector $e_6$,
\begin{equation}
\label{e4.7}
 t^{(\Om)}_a =  \sum_{k=1}^{27} (\om_k \cdot y)^a, \qquad
 \om_k \in \Om(e_6)\ ,
\end{equation}
(cf. (\ref{orbit})), where $a = 2,5,6,8,9,12$ are the degrees of
the $E_6$ Weyl group  and $\om_k, k=1,2,\ldots 27$ are the orbit
elements. The orbit variables $y$ in (\ref{e4.7}) are identified
with variables $y_1 \ldots y_6$ in (\ref{e4.3}) and in
(\ref{e4.7}). The Weyl-invariant polynomials of the fixed degree
are defined ambiguously
\begin{eqnarray}
\label{e4.8}
 t^{(\Om)}_2 & \rar & t^{(\Om)}_2\ ,\non \\
 t^{(\Om)}_5 & \rar & t^{(\Om)}_5\ ,\non \\
 t^{(\Om)}_6 & \rar & t^{(\Om)}_6 + A^{(6)} (t^{(\Om)}_2)^3 \ ,\non \\
 t^{(\Om)}_8 & \rar & t^{(\Om)}_8 + A^{(8)}_1 t^{(\Om)}_2 t^{(\Om)}_6
 + A^{(8)}_2 (t^{(\Om)}_2)^4\ ,\non \\
 t^{(\Om)}_9 & \rar & t^{(\Om)}_9 + A^{(9)} (t^{(\Om)}_2)^2 t^{(\Om)}_5\ ,\\
 t^{(\Om)}_{12} & \rar & t^{(\Om)}_{12} + A^{(12)}_1 t^{(\Om)}_2 (t^{(\Om)}_5)^2
 + A^{(12)}_2 (t^{(\Om)}_2)^2 t^{(\Om)}_8 +
 A^{(12)}_3 (t^{(\Om)}_2)^3 t^{(\Om)}_6 \non \\
&& + A^{(12)}_4 (t^{(\Om)}_2)^6 + A^{(12)}_5 (t^{(\Om)}_6)^2\
.\non
\end{eqnarray}
where $A^{(6,8,9,12)}_i$ are parameters. For general values of these
parameters one gets the algebraic Hamiltonian preserving the flag
$2,5,6,8,9,12$ and even smaller one $1,2,3,4,4,6$. Our goal, as
before, is to tune these parameters in such a way that the Hamiltonian
in its algebraic form preserves the minimal flag. Quite cumbersome
analysis results to a one-parametric set of variables
\begin{eqnarray}
\label{e4.9}
 \ta_2 &=&  \frac{4}{3}\,t^{(\Om)}_2\ , \non \\
 \ta_5 &=&  \frac{576}{5} \,t^{(\Om)}_5 \ ,\non \\
 \ta_6 &=& 3456\,t^{(\Om)}_6 - 24\,(t^{(\Om)}_2)^3\ , \non \\
 \ta_8 &=&  \frac{248832}{5} \,t^{(\Om)}_8
+ 48\,(t^{(\Om)}_2)^4 -  \frac {55296}{5} \,t^{(\Om)}_2\,t^{(\Om)}_6 \ ,\\
 \ta_9 &=& { \frac {663552\,{t^{(\Om)}_9}}{7}}  -
{ \frac {27648\,{t^{(\Om)}_5}\,{t^{(\Om)}_2}^{2}}{5}}\ , \non \\
 \ta_{12} &=&  \frac {95551488}{5}
\,t^{(\Om)}_{12}  -
 \frac{5568}{5}\,(t^{(\Om)}_2)^6  + 294912\,t^{(\Om)}_6\,(t^{(\Om)}_2)^3
 + A_1^{(12)} t^{(\Om)}_2\,(t^{(\Om)}_5)^2\non
 \\
&& - 1658880\,(t^{(\Om)}_2)^2\,t^{(\Om)}_8 - \frac{5308416}{5}
\,(t^{(\Om)}_6)^2 \ ,\non
\end{eqnarray}
where $A_1^{(12)}$ is a parameter, leading to the minimal flag, which
we denote ${\cal P}^{(E_6)}$. This flag is spanned by the spaces 
\begin{equation}
\label{e4.10}
 {\cal P}_{n}^{(1,1,2,2,2,3)} \ = \
\langle  \ta_2^{p_2} \ta_5^{p_5} \ta_6^{p_6}
\ta_8^{p_8}\ta_9^{p_9} \ta_{12}^{p_{12}} | \ 0 \leq p_2 + p_5 + 2
p_6 + 2 p_8 + 2 p_9 + 3 p_{12}\leq n \rangle\ ,
\end{equation}
with the characteristic vector,
\begin{equation}
\label{e4.11f}
 \vec f \ =\ (1,1,2,2,2,3)\ .
\end{equation}
(cf.(\ref{e2.6f}) and (\ref{e3.11f})), which coincides with the $E_6$
highest root, confirming Kac's conjecture.


The most general polynomial transformation which preserves a linear
space ${\cal P}_{n}^{(E_6)}={\cal P}^{(1,1,2,2,2,3)}_n$ for any
$n$ is of the form
\[
  s_2 \rar s_2 + a_2 s_5\ ,
\]
\[
 s_5 \rar s_5 + a_5 s_2\ ,
\]
\[
 s_6 \rar s_6 + a_{6,1} s_2^2 + a_{6,2} s_5^2+b_{6,1} s_8 +
b_{6,2} s_9+ b_{6,3} s_2 s_5 \ ,
\]
\[
 s_8 \rar s_8 + a_{8,1} s_2^2 + a_{8,2} s_5^2+b_{8,1} s_6 +
b_{8,2} s_9+ b_{8,3} s_2 s_5 \ ,
\]
\[
 s_9 \rar s_9 + a_{8,1} s_2^2 + a_{8,2} s_5^2+b_{8,1} s_6 +
b_{8,2} s_8+ b_{8,3} s_2 s_5 \ ,
\]
\[
 s_{12} \rar s_{12} + a_{12,1} s_2^3 + a_{12,2} s_5^3 +
b_{12,1} s_2^2 s_5 +b_{12,2} s_2 s_5^2 + c_{12,1} s_2 s_6+
c_{12,2} s_2 s_8
\]
\begin{equation}
\label{e4.12t}
 +c_{12,3} s_2 s_9 +d_{12,1} s_5 s_6+d_{12,2} s_5
s_8+d_{12,3} s_5 s_9 \ ,
\end{equation}
where $\{a,b,c,d\}$ are arbitrary numbers. Surprisingly, there is an
overlap of non-linear transformations (\ref{e4.9}) and (\ref{e4.12t}).
Namely, a variation of the parameter $A^{(12)}_1$ in (\ref{e4.9})
corresponds to varying the parameter $b_{12,2}$ in the transformation
(\ref{e4.12t}). Therefore there exists a non-trivial one-parametric
set of invariants of fixed degrees leading to an algebraic form of the
Hamiltonian $h_{E_6}^{(r)}$ and simultaneously preserving a minimal
flag !  Thus, the parameter $A^{(12)}_1$ can be chosen following our
convenience. We set $A^{(12)}_1=0$, which makes the coefficient
functions in the algebraic form of the Hamiltonian (see below
(\ref{e4.13})) the polynomials of lowest degree.  However, it is still
an open question for which value(s) of this parameter the invariants
(\ref{e4.9}) can be `trigonometrized' leading to an algebraic form of the
trigonometric $E_6$ model.

Finally, the rational Hamiltonian $h_{E_6}^{(r)} (y_1\ldots y_6)$
can be written as
\begin{equation}
\label{e4.13}
 h_{E_6}^{(r)} (y_1\ldots y_6) =  {\cal A}_{a,b}
 \frac{\pa^2\ }{\pa \ta_a \pa \ta_b}  +
 {\cal B}_a \frac{\pa}{\pa \ta_a}\ ,
\end{equation}
where summation over $a,b = 2,5,6,8,9,12$ with $a\leq b$ is
carried out with the coefficient functions:
\[
{\cal A}_{2,2}=8 \ta_2\ ,\ {\cal A}_{2,5}=20 \ta_5\ , \ {\cal
A}_{2,6}= 24 \ta_6\ ,
\]
\[
{\cal A}_{2,8}= 32 \ta_8\ ,\ {\cal A}_{2,9}=36 \ta_9\ ,\ {\cal
A}_{2,12}= 48 \ta_{12}\ ,
\]
\[
{\cal A}_{5,5}=4 \ta_8\ ,\ {\cal A}_{5,6}=54 \ta_9 + 243 {\ta_2}^2
\ta_5\ ,\ {\cal A}_{5,8}=48 \ta_5 \ta_6 + 90 \ta_2 \ta_9\ ,
\]
\[
\ {\cal A}_{5,9}=12 \ta_{12} + 162 \ta_2 {\ta_5}^2\ ,\ {\cal
A}_{5,12}= 36 \ta_6 \ta_9 + 81\ta_2 \ta_5 \ta_8\ ,
\]
\[
{\cal A}_{6,6}=1080{\ta_5}^2 + 270\ta_2 \ta_8 + 162{\ta_2}^2
\ta_6\ ,\ {\cal A}_{6,8}=144 \ta_{12} + 2754\ta_2 \ta_5^2 + 324
\ta_2^2 \ta_8 \ ,
\]
\[
{\cal A}_{6,9}=234 \ta_5\ta_8 + 405 {\ta_2}^2 \ta_9\ ,
\ {\cal A}_{6,12} = 567 \ta_2 \ta_5 \ta_9 + 72{\ta_8}^2 +
540{\ta_5}^2 \ta_6 + 486 {\ta_2}^2 \ta_{12}\ ,
\]
\[
{\cal A}_{8,8} = 4374 \ta_2^2 \ta_5^2 + 48\ta_6\ta_8 +
504\ta_5\ta_9 + 216 \ta_2 \ta_{12}\ ,
\]
\[
{\cal A}_{8,9}=540 \ta_5^3 + 72\ta_6\ta_9 + 378 \ta_2 \ta_5 \ta_8\
,
\]
\[
 {\cal A}_{8,12} = 729 \ta_2^2 \ta_5
 \ta_9 + 162{\ta_9}^2 + 72\ta_6 \ta_{12} + 270{\ta_5}^2\ta_8
 + 108 \ta_2 \ta_8^2 + 972 \ta_2 \ta_5^2 \ta_6\ ,
\]
\[
{\cal A}_{9,9} = 216 \ta_2\ta_5\ta_9 + 12\ta_8^2 +
144\ta_5^2\ta_6\ ,
\]
\[
 {\cal A}_{9,12} = -162 \ta_2 \ta_5 \ta_{12} + 270\ta_5^2\ta_9
 + 189 \ta_2\ta_8\ta_9 + 144 \ta_5\ta_6\ta_8\ ,
\]
\[
 {\cal A}_{12,12} = \frac{1215}{2} \ta_2^2\ta_9^2 - 324\ta_2\ta_5\ta_6\ta_9
 +36\ta_6\ta_8^2-648\ta_5^2\ta_{12}+432\ta_5^2\ta_6^2\non
\]
\begin{equation}
\label{e4.14}
 + 594 \ta_5\ta_8\ta_9  + 162 \ta_2\ta_8\ta_{12}\ ,
\end{equation}
and
\[
 {\cal B}_{2}= - 4\om \ta_2 + 24 (1+12\nu)\ ,\
 {\cal B}_{5}= - 10 \om \ta_5\ ,\
 {\cal B}_{6}= - 12\om \ta_6 + 405 (1+6\nu) {\ta_2}^2\ ,
\]
\[
{\cal B}_{8}= - 16\om \ta_8 + 96(1+3\nu)\ta_6\ ,\ {\cal B}_{9}= -
18 \om \ta_9 + 216(2+3\nu)\ta_2 \ta_5\ ,
\]
\begin{equation}
\label{e4.15}
 {\cal B}_{12}\ =\ - 24 \om \ta_{12} + 108(7-6\nu)\ta_5^2 +
 324(1 + 2\nu) \ta_2 \ta_8\ .
\end{equation}

There is a one-parametric algebra of differential operators (in six
variables) for which ${\cal P}^{(1,1,2,2,2,3)}_n$ (see (\ref{e4.10}))
is a finite-dimensional irreducible representation space. Furthermore,
the finite-dimensional representation spaces appear for different
integer values of the algebra parameter. They form an infinite
non-classical flag which coincides with ${\cal P}^{(E_6)}$
(\ref{e4.10}). We call this algebra $e^{(6)}$.  Like the algebras
$g^{(2)}, f^{(4)}$ introduced in \cite{Rosenbaum:1998}) and
\cite{blt}, respectively, in relation to the $G_2$ and $F_4$ models,
the algebra $e^{(6)}$ is infinite-dimensional yet finitely-generated.
It will be described and studied elsewhere. The rational $E_6$
Hamiltonian in the algebraic form (\ref{e4.13}) with coefficients
(\ref{e4.14}), (\ref{e4.15}) can be rewritten in terms of the
generators of this algebra.

The operator (\ref{e4.13}) is triangular in the basis of monomials
$\ta_2^{p_1}\ta_5^{p_2}\ta_6^{p_3}\ta_8^{p_4}\ta_9^{p_5}\ta_{12}^{p_6}$.
One can find the spectrum of (\ref{e4.13}), $h_{\rm E_6}^{(r)} \varphi
= -2\ep\varphi$, explicitly
\begin{equation}
\label{e4.16}
  \ep_{n_1,n_2,n_3,n_4,n_5,n_6}= \om (2n_1 + 5 n_2 + 6 n_3+ 8 n_4+9 n_5+ 12 n_6) \ ,
\end{equation}
where $n_i$ are non-negative integers. Degeneracy of the spectrum is
related to the number of partitions of an integer number $n,\ 
n=0,1,2,\ldots$ to $2n_1 + 5 n_2 + 6 n_3+ 8 n_4+9 n_5+ 12 n_6$.  The
spectrum does not depend on the coupling constant $g$, it is
equidistant and corresponds to the spectrum of a set of the harmonic
oscillators. Finally, the energies of the original rational $E_6$
Hamiltonian (\ref{e4.1}) are $E=E_0+\ep$. As an illustration the first
eigenfunctions are presented in the Appendix B. It is worth noting
that the Hamiltonian (\ref{e4.13}) possesses a remarkable property:
there exists a family of eigenfunctions which depend on the single
variable $\ta_2$. These eigenfunctions are the associated Laguerre
polynomials. This property admits to construct a
quasi-exactly-solvable generalization of the rational $E_6$ model, as
will be done elsewhere. Due to enormous technical difficulties we were
unable to describe explicitly the boundary of the configuration space
in the Weyl-invariant variables $\ta$'s similarly to what was done in
$G_2$ and $F_4$ cases.


\setcounter{equation}{0}
\section{The rational $E_7$ model}

The Hamiltonian of the rational $E_7$ model is built using the root
system of the exceptional $E_7$ algebra. A convenient way to represent
the Hamiltonian is to write it in the $8-$dimensional space
$\{x_1,x_2,\ldots x_8\}$ and impose the constraint $x_8=-x_7$,
\begin{equation}
  \label{e5.1}
  H_{E_7} = -\frac{1}{2} \Delta^{(8)}
  + \frac{\om^2}{2} \ \sum_{i=1}^{8} x_i^2 +V_{E_7}\ ,
\end{equation}
where $\om$ is a frequency and the root generated part of the
potential depends on a single constant $g=\nu(\nu-1)$:
\begin{align}
\label{e5.2}
 V_{E_7} =&  g \, \sum_{j<i =1}^{6}
 \left[
\frac{1}{(x_i +x_j)^2} + \frac{1}{(x_i - x_j)^2}
 \right] + g \frac{1}{(x_7 - x_8)^2}
\non \\
  &~~~~~~~~~~~~+ g \sum^6_{\nu_j} \frac{1}{\left[\frac{1}{2}\left({
-x_8 + x_7  - \sum_{j=1}^{6} (-1)^{\nu_j}x_j }\right)\right]^2}\ ,
\end{align}
with $\nu_j=0,1$, and $\sum_{j=1}^6 \nu_j =$ odd. The
configuration space is given by the principal $E_7$ Weyl chamber.

Let us introduce the new variables
\begin{eqnarray*}
 y_i &=& x_i\ ,\quad i=1\ldots 6 \\
 y_7 &=& x_7-x_8\ ,\qquad \mbox{ (using the constraint $y_7=2x_7$)}
\\
 Y &=& \frac{1}{2} (x_7 + x_8)\ ,\qquad \mbox{(using the constraint
 $Y=0$)}
\end{eqnarray*}
In these variables the Laplacian becomes
\begin{equation}
\label{e5.3}
 \De^{(8)} = \De_y^{(6)} + 2 \frac{\pa^2}{\pa y_7^2} +
\frac{1}{2} \frac{\pa^2}{\pa Y^2}\ ,
\end{equation}
and the potential part of (\ref{e5.1}) depends only on $\{y_1
\ldots y_7 \}$:
\begin{eqnarray}
\label{e5.4}
 V &=& \frac{\om^2}{2} \left\{
 \sum_{i=1}^{6} y_i^2 + \frac{y_7^2}{2} \right\} + g \sum_{j<i =1}^{6}
  \left[
\frac{1}{(y_i + y_j)^2} + \frac{1}{(y_i - y_j)^2}
  \right] + \frac{g}{{y_7}^2}
\non \\ &&
 + g \sum^6_{\nu_j} \frac{1}{\left[\frac{1}{2}\left({y_7
- \sum_{j=1}^{6} (-1)^{\nu_j}y_j }\right)\right]^2}\ .
\end{eqnarray}
In this formalism imposing constraints means the restriction to the
eigenfunctions having no dependence on $Y$. Hence, $Y$-dependent part
of the Laplacian, the last term in (\ref{e5.3}), can be dropped off.

The ground state eigenfunction has a form
\begin{equation}
\label{e5.6}
 \Psi_0 = (\De_+^{(6)} \De_-^{(6)}y_7)^\nu \De_{E_7}^\nu
 {\rm e}^{-\frac{1}{2} \om \left\{
 \sum_{i=1}^{6} y_i^2 + \frac{y_7^2}{2} \right\}} \ ,\
 E_0 = \frac{7}{2} \om (1+18\nu )\ ,
\end{equation}
where
\begin{eqnarray*}
 \De_\pm^{(6)} &=& \prod_{j<i =1}^{6} (y_i \pm y_j)\ , \\
 \De_{E_7} &=& \prod_{\{\nu_j\}}
 \left(y_7 + \sum_{j=1}^{6} (-1)^{\nu_j}y_j \right)\ ,
\end{eqnarray*}
with $\nu_j=0,1$ and  $\sum_{j=1}^{6} \nu_j = $ odd and $g = \nu
(\nu-1)$.

 In order to find variables leading to algebraic form of
gauge-rotated Hamiltonian,
\begin{equation}
\label{e5.6h}
 h_{E_7}^{(r)} (y_1\ldots y_7)\ =\ -2{\Psi_0}^{-1}
 ({\cal H}_{E_7}-E_0)(y_1\ldots y_7){\Psi_0}\ ,
\end{equation}
let us take the Weyl-invariant polynomials, obtained by averaging
over the 56-dimensional orbit $\Om$ generated by the vector
$(e_7-e_6)$,
\begin{equation}
\label{e5.7}
 t^{(\Om)}_a =  \sum_{k=1}^{56} (\om_k \cdot x)^a, \qquad
 \om_k \in \Om(e_7-e_6)\ ,
\end{equation}
(cf. (\ref{orbit})), where $a=2,6,8,10,12,14,18$ are the degrees
of the $E_7$ invariants and $\om_k, k=1,2,\ldots 56$ are the orbit
elements. The orbit variables $t^{(\Om)}_a$ are functions of $y_1
\ldots y_7$. The invariants of a fixed degree are defined
ambiguously, up to non-linear transformations, similar to
(\ref{e2.8t}), (\ref{e3.9t}), (\ref{e4.8}) 
\begin{eqnarray}
\label{e5.8}
 t^{(\Om)}_2 & \mapsto & t^{(\Om)}_2\ ,\non \\
 t^{(\Om)}_6 & \mapsto & t^{(\Om)}_6 + A^{(6)} (t^{(\Om)}_2)^3\ ,\non \\
 t^{(\Om)}_8 & \mapsto & t^{(\Om)}_8 + A^{(8)}_1 t^{(\Om)}_2 t^{(\Om)}_6
 + A^{(8)}_2 (t^{(\Om)}_2)^4\ ,\non \\
 t^{(\Om)}_{10} & \mapsto & t^{(\Om)}_{10} + A^{(10)}_1 (t^{(\Om)}_2)^2
 t^{(\Om)}_6 + A^{(10)}_2 t^{(\Om)}_2
 t^{(\Om)}_8 + A^{(10)}_3 (t^{(\Om)}_2)^5\ ,\\
 t^{(\Om)}_{12} & \mapsto & t^{(\Om)}_{12} + A^{(12)}_1 (t^{(\Om)}_6)^2
 + A^{(12)}_2 t^{(\Om)}_2 t^{(\Om)}_{10}
 + A^{(12)}_3 (t^{(\Om)}_2)^2 t^{(\Om)}_8 +
 A^{(12)}_4 (t^{(\Om)}_2)^3 t^{(\Om)}_6 \non \\
 && + A^{(12)}_5 (t^{(\Om)}_2)^6\ ,\non
\\
 t^{(\Om)}_{14} & \mapsto & t^{(\Om)}_{14} + A^{(14)}_1 t^{(\Om)}_2
 (t^{(\Om)}_6)^2
 + A^{(14)}_2 t^{(\Om)}_2 t^{(\Om)}_{12}
 + A^{(14)}_3 (t^{(\Om)}_2)^2 t^{(\Om)}_{10}
 + A^{(14)}_4 (t^{(\Om)}_2)^3 t^{(\Om)}_8 \non \\
&& + A^{(14)}_5 t^{(\Om)}_6 t^{(\Om)}_8 + A^{(14)}_6
(t^{(\Om)}_2)^4 t^{(\Om)}_6 + A^{(14)}_7 (t^{(\Om)}_2)^7\ ,\non
\\
 t^{(\Om)}_{18} & \mapsto & t^{(\Om)}_{18} + A^{(18)}_1 (t^{(\Om)}_6)^3
 + A^{(18)}_2 t^{(\Om)}_6 t^{(\Om)}_{12}
 + A^{(18)}_3 t^{(\Om)}_8 t^{(\Om)}_{10}
 + A^{(18)}_4 t^{(\Om)}_2 (t^{(\Om)}_{8})^2 \non
\\
&&
 + A^{(18)}_5 t^{(\Om)}_2 t^{(\Om)}_6 t^{(\Om)}_{10}
 + A^{(18)}_6 (t^{(\Om)}_2)^2 t^{(\Om)}_{14}
 + A^{(18)}_7 (t^{(\Om)}_2)^2 t^{(\Om)}_{6}t^{(\Om)}_{8}
 + A^{(18)}_8 (t^{(\Om)}_2)^3 t^{(\Om)}_{12} \non \\
&&
 + A^{(18)}_9 (t^{(\Om)}_2)^3 (t^{(\Om)}_{6})^2
 + A^{(18)}_{10} (t^{(\Om)}_2)^4 t^{(\Om)}_{10}
 + A^{(18)}_{11} (t^{(\Om)}_2)^5 t^{(\Om)}_{8}
 + A^{(18)}_{12} (t^{(\Om)}_2)^5 t^{(\Om)}_{6}\non \\&&
 + A^{(18)}_{13} (t^{(\Om)}_2)^9\ .\non
\end{eqnarray}
Our goal is to tune the parameters $A$'s so as to get the algebraic
form of the Hamiltonian (if it exists) and a minimal flag. After very
cumbersome analysis we discovered a two-parametric set of variables
(for simplicity we omit subscript $(\Om)$ in variables $t^{(\Om)}_a$),
\begin{align}
\label{e5.9}
 \ta_2 =& \frac{1}{3} t_2\ ,\
 \ta_6 =  -\frac{16}{3} t_6 + \frac{1}{108} t_2^3\ ,\
 \ta_8 = -\frac{16}{5}t_8 + \frac{16}{45}t_2 t_6-\frac{1}{2592} t_2^4\ ,
  \non \\[5pt]
 \ta_{10} =& \frac{64}{315}t_{10} - \frac{4}{105}t_2 t_8
-\frac{1}{466560}t_2^5 + \frac{1}{405} t_2^2 t_6 \ , \non \\[5pt]
 \ta_{12} =& \frac {1024}{45}\, t_{12} + \frac{5}{419904} t_2^6 -
\frac{64}{3645} t_2^3 t_6 + \frac{184}{405} t_2^2 t_8 -
A^{(12)}_2 t_2 t_{10} - \frac{256}{405} t_6^2\ , \non \\[5pt]
 \ta_{14} =& \frac{4096}{2233} t_{14}  - \frac{43}{634230} t_2^4
t_6 + \frac{832}{35235} t_2 t_6^2 - \frac{256}{1305} t_6 t_8 -
\frac{202}{246645} t_2^3 t_8 \non
\\[5pt] &  +
\frac{341}{5114430720}t_2^7 - \frac{18176}{43065}t_2t_{12} +
\frac{8768}{246645} t_2^2 t_{10}\ , \non \\[5pt]
 \ta_{18} =&
\frac{262144}{3687} t_{18} + \frac{635120768}{505211175} t_2^2 t_6
t_8 - \frac{534492928}{5846015025} t_2^3 t_6^2 +
\frac{49527971}{210456540900} t_6 t_2^6 \non
\\[5pt]
& - \frac{4002704}{5846015025} t_2^5 t_8 +
\frac{192754688}{285805179} t_2^3 t_{12} -
\frac{9332528}{233840601} t_2^4 t_{10} -
\frac{2735503}{16971215458176} t_2^9 \non
\\[5pt]
& + \frac{262144}{1493235} t_6^3 - \frac{54099968}{17421075} t_2
t_6 t_{10} - A^{(18)}_3 t_8 t_{10} - \frac{38912}{30725} t_2 t_8^2
\non
\\[5pt]
& - \frac{458752}{55305} t_6 t_{12} - \frac{98099200}{24699213}
t_2^2 t_{14}\ ,
\end{align}
where $A^{(12)}_2, A^{(18)}_3$ are parameters, leading to an
algebraic form of the Hamiltonian and in which the flag is
minimal. We denote the minimal flag ${\cal P}^{(E_7)}$. This flag
is generated by the polynomials
\pagebreak
\begin{align}
\label{e5.10}
 {\cal P}_{n}^{(1,2,2,2,3,3,4)} =
\langle  \ta_2^{p_2} \ta_6^{p_6} \ta_8^{p_8} \ta_{10}^{p_{10}}
\ta_{12}^{p_{12}} \ta_{14}^{p_{14}}\ta_{18}^{p_{18}} |& \ 0 \leq
p_2 +
2p_6 + 2 p_8 \non\\
& + 2 p_{10} + 3 p_{12} + 3 p_{14}+ 4 p_{18}\leq n \rangle\ ,
\end{align}\nopagebreak
with the characteristic vector
\begin{equation}
\label{e5.11f}
 \vec f \ =\ (1,2,2,2,3,3,4)\ ,
\end{equation}
(cf.(\ref{e2.6f}), (\ref{e3.11f}), (\ref{e4.11f})). It again coincides
with the highest root of the $E_7$ algebra confirming the Kac's
conjecture. Therefore ${\cal P}_{n}^{(E_7)}={\cal  P}_{n}^{(1,2,2,2,3,3,4)}$.


It is worth to emphasize that the most general polynomial
transformation which preserve each linear space ${\cal
P}_{n}^{(1,2,2,2,3,3,4)}$ is
\[
 s_2 \rar s_2 \ ,
\]
\[
 s_6 \rar s_6 + a_6 s_2^2+ b_{6,1} s_8 + b_{6,2} s_{10} \ ,
\]
\[
 s_8 \rar s_8 + a_8 s_2^2+ b_{8,1} s_6 + b_{8,2} s_{10} \ ,
\]
\[
 s_{10} \rar s_{10} + a_{10} s_2^2+ b_{10,1} s_6 + b_{10,2} s_{8} \ ,
\]
\[
 s_{12} \rar s_{12} + a_{12} s_2^3  +
b_{12,1} s_2 s_6 +b_{12,2} s_2 s_8 + b_{12,3} s_2 s_{10}+ c_{12}
s_{14}\ ,
\]
\[
 s_{14} \rar s_{14} + a_{14} s_2^3  +
b_{14,1} s_2 s_6 +b_{14,2} s_2 s_8 + b_{14,3} s_2 s_{10}+ c_{14}
s_{12}\ ,
\]
\[
 s_{18} \rar s_{18} + a_{18} s_2^4 +
b_{18,1} s_2^2 s_6 +b_{18,2} s_2^2 s_8 +b_{18,3} s_2^2 s_{10} +
c_{18,1} s_2 s_{12}+ c_{18,2} s_2 s_{14}
\]
\begin{equation}
\label{e5.12t}
 + d_{18,1} s_6^2 + d_{18,2} s_8^2 + d_{18,3} s_{10}^2
 + d_{18,4} s_6 s_8 + d_{18,4} s_6 s_{10} + d_{18,4} s_8 s_{10} \ ,
\end{equation}
where $\{a,b,c,d\}$ are arbitrary numbers. It is worth to mention that
there exist two exceptional sub-transformations which are common for
(\ref{e5.8}) and (\ref{e5.12t}) -- when we vary the parameter
$b_{12,3}$ in $s_{12}$ ($A^{(12)}_2$ in $t^{(\Om)}_{12}$) and
$d_{18,4}$ in $s_{18}$ ($A^{(18)}_3$ in $t^{(\Om)}_{18}$) keeping all
other parameters fixed. Thus, we conclude that there exist a
two-parametric set of invariants of the fixed degrees leading to
algebraic form of the Hamiltonian $h_{E_7}^{(r)}$ and moreover
preserving the minimal flag. Hence the parameters $A^{(12)}_2$ and
$A^{(18)}_3$ can be chosen at our will. We set them equal to zero,
$A^{(12)}_2=A^{(18)}_3=0$, which fixes the coefficient functions in
the algebraic form of the Hamiltonian (see below (\ref{e5.13})) in the
form of polynomials of lowest degrees. Since an algebraic form of the
trigonometric $E_7$ model is not known so far, it is an open question
for which value(s) of these parameters the invariants (\ref{e5.8})
would be `trigonometrized' leading to an algebraic form of the
trigonometric $E_7$ model (if such a form exists).

Finally, the Hamiltonian $h_{E_7}^{(r)}$ can be written as
\begin{equation}
\label{e5.13}
 h_{E_7}^{(r)}\ =\  {\cal A}_{a\,b}
 \frac{\pa^2\ }{\pa \ta_a \pa \ta_b}  +
 {\cal B}_a \frac{\pa}{\pa \ta_a}\  \quad  ~,
\end{equation}
where $a,b=2,6,8,10,12,14,18$ and $a \leq b$  with the coefficient
functions:
\begin{align}
\label{e5.14}
 {\cal A}_{2,2}\ \ &= 16 \ta_2\ ,\ {\cal A}_{2,6} =
48 \ta_6\ ,\ {\cal
A}_{2,8} = 64 \ta_8\ ,\ {\cal A}_{2,10} = 80 \ta_{10}\ , \non\\
{\cal A}_{2,12}\ &= 96 \ta_{12}\ ,\ {\cal A}_{2,14} = 112
\ta_{14}\
,\ {\cal A}_{2,18} = 144 \ta_{18}\ , \non\\
{\cal A}_{6,6}\ \ &= - 8 \ta_2^2 \ta_6 - 160 \ta_2 \ta_8 + 17280
\ta_{10}\ ,\ {\cal A}_{6,8} =  1920 \ta_2 \ta_{10} +  96 \ta_{12}
- 16 \ta_2^2 \ta_8\ , \non\\
{\cal A}_{6,10}\ &= -  112 \ta_{14}  - 24 \ta_2^2 \ta_{10}\ ,\
{\cal A}_{6,12}= - 128 \ta_8^2 + 3456 \ta_{10} \ta_6 - 1440 \ta_2
\ta_{14} - 24 \ta_2^2 \ta_{12}\ , \non\\
{\cal A}_{6,14}\ &= - 128 \ta_8 \ta_{10} - 36 \ta_{18}  - 32
\ta_2^2 \ta_{14}\ ,\ {\cal A}_{6,18}=  - 23040 \ta_2 \ta_{10}^2 -
40
\ta_2^2 \ta_{18} \non\\
& + 3200 \ta_8 \ta_{14} - 4224 \ta_{10} \ta_{12}\ , \non\\
{\cal A}_{8,8}\ \  &=  336 \ta_{14}  + 288 \ta_2^2 \ta_{10} + 8
\ta_6 \ta_8 + 12 \ta_2 \ta_{12}\ ,\ {\cal A}_{8,10} = - 16 \ta_2
\ta_{14}  + 16 \ta_6 \ta_{10} \ , \non\\
{\cal A}_{8,12}\ &= - 1344 \ta_8 \ta_{10} + 480 \ta_2 \ta_6
\ta_{10} - 108 \ta_{18} - 16 \ta_2 \ta_8^2 + 12 \ta_6 \ta_{12}  -
192
\ta_2^2 \ta_{14} \ , \non\\
{\cal A}_{8,14}\ &= - 960 \ta_{10}^2 - 16 \ta_2 \ta_8 \ta_{10} +
20\ta_6 \ta_{14} - 5\ta_2 \ta_{18}\ , \non\\
{\cal A}_{8,18}\ &= - 3456 \ta_2^2 \ta_{10}^2 - 576 \ta_2 \ta_{10}
\ta_{12} + 448 \ta_2 \ta_8 \ta_{14} + 24 \ta_6 \ta_{18} - 9984
\ta_{10} \ta_{14}\ , \non\\
{\cal A}_{10,10} &=  \frac{4 \ta_8 \ta_{10}}{3}  +
\frac{\ta_{18}}{6}\ ,\ {\cal A}_{10,12} = 960 \ta_{10}^2 - 32
\ta_2 \ta_8 \ta_{10} -  20 \ta_6 \ta_{14} + \ta_2 \ta_{18}\ , \non\\
{\cal A}_{10,14} &= - \frac {8 \ta_8\ta_{14}}{3}  +
4\ta_{10}\ta_{12}\ ,\ {\cal A}_{10,18} = 288 \ta_2 \ta_{10}
\ta_{14} - \frac{8 \ta_8 \ta_{18}}{3} + 16 \ta_{12} \ta_{14} - 384
\ta_6 \ta_{10}^2 \ ,\ \non\\
{\cal A}_{12,12} &= 11520  \ta_2 \ta_{10}^2 + 192  \ta_2^2
\ta_{10} \ta_8 - 24 \ta_2 \ta_8 \ta_{12}
 +  36 \ta_2^2 \ta_{18}  -  336 \ta_2 \ta_6 \ta_{14} \non\\
& - 16 \ta_6 \ta_8^2 - 1056 \ta_8 \ta_{14} + 1152 \ta_{10}
\ta_{12}
+ 576 \ta_6^2 \ta_{10}\ , \non\\
{\cal A}_{12,14} &= 288 \ta_2^2 \ta_{10}^2 + 24 \ta_2 \ta_{10}
\ta_{12} - 64 \ta_2 \ta_8 \ta_{14} - 16 \ta_6 \ta_8 \ta_{10} - 6
\ta_6 \ta_{18}  + 1824 \ta_{10} \ta_{14}\ , \non\\
{\cal A}_{12,18} &= 2688 \ta_{14}^2 + 73728 \ta_2^2 \ta_{10}
\ta_{14} - 672 \ta_6 \ta_{10} \ta_{12} + 544 \ta_6 \ta_8  \ta_{14}
+ 384 \ta_2 \ta_{12} \ta_{14} \non\\
& - 288 \ta_2 \ta_8 \ta_{18}  + 9216 \ta_8 \ta_{10}^2 + 2592
\ta_{10} T_{18} - 27648 \ta_2 \ta_6 \ta_{10}^2\ , \non\\
{\cal A}_{14,14} &= 32 \ta_2 \ta_{10} \ta_{14} -  \frac{2 \ta_8
\ta_{18}}{3} + 4 \ta_{12} \ta_{14}  - 64 \ta_6 \ta_{10}^2\ , \non\\
{\cal A}_{14,18} &= 23040 \ta_{10}^3 + 64 \ta_8^2 \ta_{14}  + 128
\ta_2 \ta_{14}^2 - 384 \ta_2 \ta_{10}^2 \ta_8 - 608 \ta_{10} \ta_6
\ta_{14} \non\\
& + 56 \ta_2 \ta_{10} \ta_{18}  - 64 \ta_8 \ta_{10} \ta_{12} +  4
\ta_{12} \ta_{18}\ , \non\\
{\cal A}_{18,18} &= 55296 \ta_2^2 \ta_{10}^3 - 768  \ta_6 \ta_{10}
\ta_{18} + 64 \ta_8^2 \ta_{18} + 285696 \ta_{10}^2 \ta_{14} + 4608
\ta_6 \ta_8 \ta_{10}^2 \non\\
& + 11520 \ta_{10}^2 \ta_2 \ta_{12} + 384\ta_{10} \ta_{12}^2 - 640
\ta_6 \ta_{14}^2  + 128 \ta_2 \ta_{14} \ta_{18} - 640 \ta_8
\ta_{12} \ta_{14} \non\\
& - 15360 \ta_2 \ta_8 \ta_{10} \ta_{14}\ ,
\end{align}
and
\begin{align}
\label{e5.15}
 {\cal B}_2\ =& - 4 \om \ta_2   + 56(1+18\nu)\ ,\
{\cal B}_6 = - 12
\om \ta_6 - 24(1 + 10\nu) \ta_2^2\ , \non\\
{\cal B}_8\ =& - 16 \om  \ta_8 + 20(1 + 6 \nu)\ta_6\ ,\ {\cal
B}_{10}= - 20 \om \ta_{10}  - 2(1 + 2\nu) \ta_8\ , \non\\
{\cal B}_{12}=& - 24 \om \ta_{12} + 480(11 + 18 \nu) \ta_{10}
-16(5
+ 18\nu) \ta_2 \ta_8\ , \non\\
{\cal B}_{14}=& - 28 \om \ta_{14} + 6 (1 + 2\nu)\ta_{12} + 24 ( 3
+
2\nu) \ta_2 \ta_{10}\ , \non\\
{\cal B}_{18}=& - 36 \om \ta_{18} - 32 (49 + 78\nu) \ta_{10} \ta_6
+ 32 (29 + 18 \nu) \ta_2 \ta_{14} + 64 (1 + 2 \nu)\ta_8^2\ . \non\\
\end{align}

There exists one-parametric algebra of differential operators (in
seven variables) for which ${\cal P}^{(1,2,2,2,3,3,4)}_n$ (see
(\ref{e5.10})) is a finite-dimensional irreducible representation.
Furthermore, the finite-dimensional representation spaces appear for
different integer values of the algebra parameter. They form an
infinite non-classical flag which coincides with ${\cal P}^{(E_7)}$
(\ref{e5.10}). We call this algebra $e^{(7)}$. Like the algebras
$g^{(2)}, f^{(4)}$ introduced in \cite{Rosenbaum:1998}) and
\cite{blt}, respectively, in relation to the $G_2$ and $F_4$ models,
the algebra $e^{(7)}$ is infinite-dimensional yet finitely-generated.
It will be described elsewhere. The rational $E_7$ Hamiltonian in the
algebraic form (\ref{e5.13}) with coefficients (\ref{e5.14}) ,
(\ref{e5.15}) can be rewritten in terms of the generators of this
algebra.

The operator (\ref{e5.13}) is triangular in the basis of monomials
$\ta_2^{p_2} \ta_6^{p_6} \ta_8^{p_8} \ta_{10}^{p_{10}}
\ta_{12}^{p_{12}} \ta_{14}^{p_{14}} \ta_{18}^{p_{18}} $. One can find
the spectrum of (\ref{e5.13}), $h_{\rm E_7}^{(r)} \varphi =
-2\ep\varphi$, explicitly,
\begin{equation}
\label{e5.16}
 \ep_{n_1,n_2,n_3,n_4,n_5,n_6,n_7}\ =\
 2\,\om (n_1 + 3 n_2 + 4 n_3 + 5 n_4 + 6n_5 + 7n_6 +
 9p_7)\ ,
\end{equation}
where $n_i$ are non-negative integers. Degeneracy of the spectrum is
related to the number of partitions of integer number $n,\ 
n=0,1,2,\ldots$ to $n_1+3n_2 + 4n_3+ 5n_4 + 6n_5 + 7n_6 + 9n_7$.  The
spectrum does not depend on the coupling constant $g$, it is
equidistant and corresponds to the spectrum of a set of harmonic
oscillators. Finally, the energies of the original rational $E_7$
Hamiltonian (\ref{e5.1}) are $E=E_0+\ep$. As an illustration the first
eigenfunctions are presented in the Appendix C. The Hamiltonian
(\ref{e5.13}) possesses a remarkable property: there exists a family
of eigenfunctions which depend on the single variable $\ta_2$. These
eigenfunctions are the associated Laguerre polynomials. This property
allows to construct a quasi-exactly-solvable generalization of the
rational $E_7$ model. It will be done elsewhere. Due to technical
difficulties we were unable to find explicitly the boundary of the
configuration space (in other words, the boundaries of the $E_7$ Weyl
chamber) in the Weyl-invariant variables $\ta$'s similar to what was
previously done for $G_2$ and $F_4$ cases (see Section 3 and 4,
correspondingly).


\setcounter{equation}{0}
\section{The rational $E_8$ model}

The Hamiltonian of the rational $E_8$ model is built using the root
system of the exceptional algebra $E_8$. The Hamiltonian is defined in
the $8-$dimensional space $\{x_1,x_2,\ldots x_8\}$ 
\begin{align}
  \label{e6.1}
  H_{E_8} = -\frac{1}{2} \Delta^{(8)} +
   \frac{\om^2}{2} \ \sum_{i=1}^{8} x_i^2 +V_{E_8}\ ,
\end{align}
where $\om$ is the oscillator parameter and $V_{E_8}$ is the
root-generated potential with the coupling constant $g=\nu(\nu-1)$:
\begin{align}
\label{e6.2}
 V_{E_8} =
 g \sum_{j<i =1}^{8}
  \left[
 \frac{1}{(x_i +x_j)^2} + \frac{1}{(x_i - x_j)^2}
  \right] + g \sum_{\nu_j}
\frac{1}{\left[\frac{1}{2}\left({ x_8 + \sum_{j=1}^{7}
(-1)^{\nu_j}x_j }\right)\right]^2}\ ,
\end{align}
where  $\nu_j=0,1$ and $\sum_{j=1}^{7} \nu_j =$ even. The
configuration space is given by the principal
$E_8$ Weyl chamber.

The ground state eigenfunction of the Hamiltonian (\ref{e6.1}) is of
the form
\begin{equation}
\label{e6.3}
 \Psi_0 = (\De_+^{(8)} \De_-^{(8)})^\nu \De_{E_8}^\nu
 {\rm e}^{-\frac{1}{2} \om
 \sum_{i=1}^{8} x_i^2 } \quad ,\quad E_0 = 4\om (1+30\nu)\ ,
\end{equation}
where
\begin{eqnarray*}
 \De_\pm^{(8)} &=& \prod_{j<i =1}^{8} (x_i \pm x_j)\ , \\
 \De_{E_8} &=& \prod_{\{\nu_j\}}
 \left(x_8 + \sum_{j=1}^{7} (-1)^{\nu_j}x_j \right)\ ,
\end{eqnarray*}
with $\nu_j=0,1\ ,\ \sum_{j=1}^{7} \nu_j = $ even and $g = \nu
(\nu-1)$.

In order to find variables leading to algebraic form of
gauge-rotated Hamiltonian,
\begin{equation}
\label{e6.3h}
 h_{E_8}^{(r)} (x_1\ldots x_8)\ =\ -2{\Psi_0}^{-1}
 ({\cal H}_{E_8} - E_0)(x_1\ldots x_8){\Psi_0}\ ,
\end{equation}
let us define a basis in the form of Weyl-invariant polynomials,
averaged over one of the smallest orbit, of length 240, generated by 
some positive root,
\begin{equation}
\label{e6.4}
 t^{(\Om)}_a =  \sum_{k=1}^{240} (\om_k \cdot x)^a, \qquad
 \om_k \in \Om(\mbox{a positive root} )\ ,
\end{equation}
(cf. (\ref{orbit})), where $a=2,8,12,14,18,20,24,30$ are degrees
of the lowest $E_8$ invariants and $\om_k, k=1,2,\ldots 240$ are
the orbit elements. The orbit variables $t^{(\Om)}_a$ are
functions of $x_1 \ldots x_8$. In general, the invariants of fixed
degree are defined ambiguously, up to a certain non-linear
transformation, cf. (\ref{e2.8t}), (\ref{e3.9t}), (\ref{e4.8}),
(\ref{e5.8}) 
\begin{eqnarray*}
    t^{(\Om)}_2 & \rar & t^{(\Om)}_2\ ,
\\[5pt]
    t^{(\Om)}_8 & \rar & t^{(\Om)}_8 +
A^{(8)} (t^{(\Om)}_2)^4\ ,
\\[5pt]
   t^{(\Om)}_{12} & \rar & t^{(\Om)}_{12} +
A^{(12)}_1 (t^{(\Om)}_2)^2 t^{(\Om)}_8 + A^{(12)}_2
(t^{(\Om)}_2)^6\ ,
\\[5pt]
   t^{(\Om)}_{14} & \rar & t^{(\Om)}_{14} +
A^{(14)}_1 t^{(\Om)}_2 t^{(\Om)}_{12} + A^{(14)}_2 (t^{(\Om)}_2)^3
t^{(\Om)}_8 + A^{(14)}_3 (t^{(\Om)}_2)^7\ ,
\\[5pt]
   t^{(\Om)}_{18} & \rar & t^{(\Om)}_{18} + A^{(18)}_1 t^{(\Om)}_2
(t^{(\Om)}_8)^2 + A^{(18)}_2 (t^{(\Om)}_2)^2 t^{(\Om)}_{14} +
A^{(18)}_3 (t^{(\Om)}_2)^3 t^{(\Om)}_{12} + A^{(18)}_4
(t^{(\Om)}_2)^5 t^{(\Om)}_8
\\&&
+ A^{(18)}_5 (t^{(\Om)}_2)^9\ ,
\\[5pt]
   t^{(\Om)}_{20} & \rar & t^{(\Om)}_{20} + A^{(20)}_1 t^{(\Om)}_8
t^{(\Om)}_{12} + A^{(20)}_2 t^{(\Om)}_2 t^{(\Om)}_{18} +
A^{(20)}_3 (t^{(\Om)}_2)^2 (t^{(\Om)}_8)^2 + A^{(20)}_4
(t^{(\Om)}_2)^3 t^{(\Om)}_{14}
\\&&
+ A^{(20)}_5 (t^{(\Om)}_2)^4 t^{(\Om)}_{12} + A^{(20)}_6
(t^{(\Om)}_2)^6 t^{(\Om)}_8 + A^{(20)}_7 (t^{(\Om)}_2)^{10}\ ,
\\[5pt]
   t^{(\Om)}_{24} & \rar & t^{(\Om)}_{24} +
A^{(24)}_1 (t^{(\Om)}_{12})^2 + A^{(24)}_2 (t^{(\Om)}_8)^3 +
A^{(24)}_3 t^{(\Om)}_2 t^{(\Om)}_8 t^{(\Om)}_{14} + A^{(24)}_4
(t^{(\Om)}_2)^2 t^{(\Om)}_{20}
 \\&&
+ A^{(24)}_5 (t^{(\Om)}_2)^2 t^{(\Om)}_8 t^{(\Om)}_{12} +
A^{(24)}_6 (t^{(\Om)}_2)^3 t^{(\Om)}_{18} + A^{(24)}_7
(t^{(\Om)}_2)^4 (t^{(\Om)}_8)^2 + A^{(24)}_8 (t^{(\Om)}_2)^5
t^{(\Om)}_{14}
 \\&&
+ A^{(24)}_9 (t^{(\Om)}_2)^6 t^{(\Om)}_{12} + A^{(24)}_{10}
(t^{(\Om)}_2)^8 t^{(\Om)}_8 + A^{(24)}_{11} (t^{(\Om)}_2)^{12}\ ,
\end{eqnarray*}
\begin{eqnarray}
\label{e6.5}
   t^{(\Om)}_{30} & \rar & t^{(\Om)}_{30} + A^{(30)}_1 t^{(\Om)}_{12}
t^{(\Om)}_{18} + A^{(30)}_2 (t^{(\Om)}_8)^2 t^{(\Om)}_{14} +
A^{(30)}_3 t^{(\Om)}_2 (t^{(\Om)}_{14})^2 + A^{(30)}_4 t^{(\Om)}_2
t^{(\Om)}_8 t^{(\Om)}_{20} \non
 \\ &&
+ A^{(30)}_5 t^{(\Om)}_2 {t^{(\Om)}_8}^2t^{(\Om)}_{12} +
A^{(30)}_6 (t^{(\Om)}_2)^2 t^{(\Om)}_{12} t^{(\Om)}_{14} +
A^{(30)}_7 (t^{(\Om)}_2)^2 t^{(\Om)}_8 t^{(\Om)}_{18} + A^{(30)}_8
(t^{(\Om)}_2)^3 t^{(\Om)}_{24} \non
\\ &&
+ A^{(30)}_9 (t^{(\Om)}_2)^3 (t^{(\Om)}_{12})^2 +
 A^{(30)}_{10} (t^{(\Om)}_2)^3 (t^{(\Om)}_8)^3 + A^{(30)}_{11}
(t^{(\Om)}_2)^4 t^{(\Om)}_8 t^{(\Om)}_{14} + A^{(30)}_{12}
(t^{(\Om)}_2)^5 t^{(\Om)}_{20} \non
 \\ &&
+ A^{(30)}_{13} (t^{(\Om)}_2)^5 t^{(\Om)}_8 t^{(\Om)}_{12} +
A^{(30)}_{14} (t^{(\Om)}_2)^6 t^{(\Om)}_{18} + A^{(30)}_{15}
(t^{(\Om)}_2)^7 (t^{(\Om)}_8)^2 + A^{(30)}_{16} (t^{(\Om)}_2)^8
t^{(\Om)}_{14} \non
 \\ &&
+ A^{(30)}_{17} (t^{(\Om)}_2)^9 t^{(\Om)}_{12} + A^{(30)}_{18}
(t^{(\Om)}_2)^{11} t^{(\Om)}_8 + A^{(30)}_{19} (t^{(\Om)}_2)^{15}\
.
\end{eqnarray}

The transformation (\ref{e6.5}) depends on $48$ parameters. Our goal
is to find parameters $A$'s such that (i) the Hamiltonian has the
algebraic form, (ii) a set of polynomial invariant subspaces forming a
flag occurs and (iii) the flag is minimal.  After extremely tedious
and cumbersome analysis we discovered a nine-parametric set of
variables (see discussion below), for simplicity we omit the subscript
$(\Om)$ in variables $t^{(\Om)}_a$,
\begin{align}
\label{e6.10}
    \ta_2\ =& \frac {1}{15} t_{2} ~, \non\\
 \ta_8\ =& \frac {1}{30} t_{8}  - \frac {13}{16200000}\,t_{2}^{4} ~,
\non\\
    \ta_{12} =& \frac {16}{21} \,t_{12} + A_1^{(12)}\,t_{2}^{2}\,t_8
+ \frac {373}{2551500000} \,t_{2}^{6} ~, \non\\
 \ta_{14} =& \frac {64}{1155} \,t_{14}  - \frac
{2531}{2296350000000} \,t_{2}^{7}  - \frac
{568}{51975}\,t_{2}\,t_{12}  + \frac {103}{1417500}
\,t_{2}^{3}\,t_{8} ~, \non\\
  \ta_{18} =& \frac {256}{4095} \,t_{18} +
 A_2^{(18)}\,t_{2}^{2}\,t_{14} +
 \frac{3706}{8353125}\,t_{2}^{3}\,t_{12}  +
 A_1^{(18)}\,t_2\,t_8^2
 - \frac{4051}{1530900000} \,t_2^5\,t_8 \non\\
&  + \frac {330961}{8266860000000000} \,t_2^9 ~, \non\\
  \ta_{20} =& \frac{4096}{575025} \,t_{20}  -
  \frac{60576512}{35320910625} \,t_2\,t_{18}
  + \frac{18641008}{779137734375}\,t_2^3\,t_{14} \non\\
&  + \frac{1}{38335}\Big( - \frac{138176}{1575}\,t_8 -
\frac{103942624}{1196015625}\,t_2^4 \Big)\,t_{12} +
\frac{323371}{6538218750}\,t_2^2\,t_8^2 \non\\
&  - \frac {10249681}{ 76261783500000000000}\,t_2^{10}  +
\frac{2994007}{353063812500000} \, t_2^6\,t_8 ~, \non\\
   \ta_{24} =& A_2^{(24)}\,t_{8}^{3}+A_3^{(24)}
\,t_2\,t_8\,t_{14}+ A_4^{(24)}\,t_2^2\,t_{20} +
\frac{32768}{13101165} \,t_{24} +
\frac{52162303808}{878607651796875}\,t_2^3 \,t_{18} \non\\
&   - \frac{2857817967448}{3663018665947265625} \,t_2^5\,t_{14}
   - \frac{197632}{362005875} \,t_{12}^{2} \non\\
&  + \frac{1}{4367055}\Big(\frac{12897642016}{31134375}
   \,t_2^2\,t_8 + \frac{11785468451047}{36777480468750}\,t_2^6\Big)
   \,t_{12} \non\\
&  - \frac{22206803851}{13661608078125000}\,t_2^4\,t_8^2
  - \frac{101271432653}{368863418109375000000} \,t_2^8\,t_8 \non\\
&  + \frac{38183226373283}{8764194814278750000000000000} \,t_2^{12} ~, \non\\
     \ta_{30} =& A_2^{(30)} \,t_8^2\,t_{14}+ A_3^{(30)}\,t_{2}\,t_{14}^{2}
+ A_4^{(30)} \,t_2\,t_8\,t_{20}+\frac{4194304}{114489375}\,t_{30} \non\\
& - \frac{87361458176}{412072580390625}  \,t_{2}^{3}\,t_{24}
  +\frac{98943157092328832}{14540238544445947265625} \,t_2^5 \,t_{20} \non\\
&+\frac{1}{7632625}\Big( - \frac{82608128}{945} \,t_{12} +
 \frac{76042100276224}{14671762875} \,t_2^2\,t_8 -
 \frac{44731593575760671656}{6158701713076171875}\,t_2^6\Big)\,t_{18} \non\\
& + \frac{1}{7632625}\Big(\frac{387292030976}{128638125}
 \,t_2^2\,t_{12}  \non\\
&  + \frac{8358520483853754832}{101437439980078125} \,t_2^4
 \,t_8 + \frac{621999713517306328312}{8558783998319091796875}
 \,t_2^8\Big)t_{14} \non\\
&  + \frac{162900154624}{81160947912515625}\,t_2^3\,t_{12}^2
  + \frac{1}{7632625}\Big(\frac{2931829717454144}{271859135625}
  \,t_2\,t_8^2 \non\\
&  - \frac{79803090068621091328}{4564684799103515625} \,t_2^5\,t_8
 - \frac{101275905787214796443}{15483227333642578125000} \,t_2^9\Big)
 t_{12} \non\\
&  - \frac{10147539336141079}{1082844671787333984375}
\,t_2^3\,t_8^3  +
  \frac{400141317989263534193}{17542083682954810546875000000}
  \,t_2^7 \,t_8^2 \non\\
&  +
\frac{105630144697706078621}{33831161388555706054687500000000}\,
 t_2^{11} \,t_8 \non\\
&  - \frac{7373766632847391460197357}
{146296767815759210806406250000000000000000} \,t_2^{15}\ ,
\end{align}
which lead to the flag which we think is minimal. We denote this
flag ${\cal P}^{(E_8)}$. The flag ${\cal P}^{(E_8)}$ is generated
by the spaces of polynomials
\begin{align}
\label{e6.10f}
 {\cal P}_{n}^{(1,3,5,5,7,7,9,11)} \ =
 \langle  \ta_2^{p_2} s_8^{p_8} \ta_{12}^{p_{12}}
\ta_{14}^{p_{14}} \ta_{18}^{p_{18}}
\ta_{20}^{p_{20}}\ta_{24}^{p_{24}} \ta_{30}^{p_{30}}|& \ 0 \leq
p_2 + 3p_8 + 5 p_{12} + 5 p_{14} + 7 p_{18} \non \\  &  + 7
p_{20}+ 9 p_{24} + 11 p_{30} \leq n \rangle\ ,
\end{align}
with the characteristic vector
\begin{equation}
\label{e6.11f}
 \vec f \ =\ (1,3,5,5,7,7,9,11) \ ,
\end{equation}
(cf.(\ref{e2.6f}), (\ref{e3.11f}), (\ref{e4.11f}), (\ref{e5.11f})).
Hence ${\cal P}^{(E_8)} = {\cal P}^{(1,3,5,5,7,7,9,11)}$. The
characteristic vector does not coincide with the highest root
$\vec{f}_{\rm highest\, root}=(2,2,3,3,4,4,5,6)$ suggested by Kac.
 We were not able to find variables in which the flag
$\vec{f}_{\rm highest\, root}$ would be preserved by the rational $E_8$
Hamiltonian.


The most general polynomial transformation which preserves each
linear space ${\cal P}_{n}^{(1,3,5,5,7,7,9,11)}, n=0,1,2,\ldots$
is
\[
 s_2 \rar s_2 \ ,
\]
\[
s_8 \rar a_{8,1}s_8 + a_{8,2}s_2^3\ ,
\]
\[
s_{12} \rar a_{12,1} s_{12} + a_{12,2} s_{14} + a_{12,3} s_2^2s_8
+ a_{12,4} s_2^5\ ,
\]
\[
 s_{14} \rar a_{14,1} s_{14} + a_{14,2} s_{12} + a_{14,3} s_2^2 s_8 + a_{14,4}
 s_2^5\ ,
\]
\[
 s_{18} \rar  a_{18,1}s_{18} + a_{18,2}
s_{20} + a_{18,3}s_2s_8^2 + a_{18,4}s_2^2s_{14} +
a_{18,5}s_2^2s_{12} + a_{18,6}s_2^4s_8  + a_{18,7}s_2^7\ ,
\]
\[
s_{20} \rar  a_{20,1} s_{20} + a_{20,2} s_{18} + a_{20,3} s_2
s_8^2 + a_{20,4} s_2^2 s_{14} + a_{20,5} s_2^2 s_{12}
 + a_{20,6} s_2^4 s_8 + a_{20,7} s_2^7\ ,
\]
\begin{eqnarray*}
s_{24} \rar  a_{24,1}s_{24} + a_{24,2}s_8^3 + a_{24,3}s_2s_8s_{14}
+ a_{24,4}s_2s_8s_{12} + a_{24,5}s_2^2s_{20}  +
a_{24,6}s_2^2s_{18}
\\
+ a_{24,7} s_2^3s_8^2 + a_{24,8}s_2^4s_{14} + a_{24,9}s_2^4s_{12}
+ a_{24,10}s_2^6s_8 + a_{24,11}s_2^9\ ,
\end{eqnarray*}
\begin{eqnarray}
\label{e6.12t}
 s_{30} \rar  a_{30,1}s_{30} + a_{30,2}s_8^2s_{14} +
a_{30,3}s_8^2s_{12} + a_{30,4}s_2s_{14}^2 + a_{30,5}s_2
s_{12}s_{14} + a_{30,6}s_2s_{12}^2 \non
\\
+ a_{30,7}s_2s_8s_{20} + a_{30,8}s_2s_8s_{18} +
a_{30,9}s_2^2s_{24} + a_{30,10}s_2^2s_8^3  +
a_{30,11}s_2^3s_8s_{14} \non
\\
 + a_{30,12}s_2^3s_8s_{12} + a_{30,13}s_2^4s_{20}
+ a_{30,14}s_2^4s_{18}  + a_{30,15}s_2^5s_8^2 +
a_{30,16}s_2^6s_{14} \non
\\
+ a_{30,17}s_2^6s_{12} + a_{30,18}s_2^8s_8 + a_{30,19}s_2^{11}\ ,
\end{eqnarray}
where $a$'s are parameters. There exist nine exceptional
sub-transformations for which (\ref{e6.5}) and (\ref{e6.12t})
coincide -- when we vary the parameter $a_{12,3}$ in $s_{12}$
($A^{(12)}_1$ in $t^{(\Om)}_{12}$) and $a_{18,3}$ and $a_{18,4}$
in $s_{18}$ ($A^{(18)}_1$ and $A^{(18)}_2$ in $t^{(\Om)}_{18}$),
correspondingly, and $a_{24,2}, a_{24,3}$ and $a_{24,5}$ in
$s_{24}$ ($A^{(24)}_2, A^{(24)}_3$ and $A^{(24)}_4$ in
$t^{(\Om)}_{24}$), correspondingly, and $a_{30,2}, a_{30,4}$ and
$a_{30,7}$ in $s_{30}$ ($A^{(30)}_2, A^{(30)}_3$ and $A^{(30)}_4$
in $t^{(\Om)}_{30}$), correspondingly. Of course, this coincidence
appears when all other parameters of the transformation are kept
fixed. Thus, one can draw a conclusion about existence of
nine-parametric set of invariants of the fixed degrees leading to
algebraic form of the Hamiltonian $h_{E_8}^{(r)}$ and moreover
preserving the flag ${\cal P}^{(1,3,5,5,7,7,9,11)}$. Hence, the
parameters can be chosen by following our convenience. In a
simply-minded way we set all of them equal to zero,
\[
A^{(12)}_1=A^{(18)}_1=A^{(18)}_2=A^{(24)}_2 =
A^{(24)}_3=A^{(24)}_3=A^{(30)}_2=A^{(30)}_3=A^{(30)}_4=0\ ,
\]
which fixes the coefficient functions in the algebraic form of the
Hamiltonian (see below (\ref{e6.13})) in form of polynomials of
lowest degrees. It is an open question for what value(s) of these
parameters the invariants (\ref{e6.5}) are `trigonometrized'
leading to an algebraic form of the trigonometric $E_8$ model, if
such a form exists.

Finally, the Hamiltonian $h_{E_8}^{(r)} (x_1\ldots x_8)$ can be
written as
\begin{align}
\label{e6.13}
 h_{E_8}^{(r)} (x_1\ldots x_8) =  {\cal A}_{a\,b}
 \frac{\pa^2\ }{\pa \ta_a \pa \ta_b}  +
 {\cal B}_a \frac{\pa}{\pa \ta_b} \qquad  ,
\end{align}
where $a,b = 2,8,12,14,18,20,24,30$ and $a\leq b$ with the
coefficient functions:
\[
{\cal A}_{2,2}=8 \ta_2\ ,\ {\cal A}_{2,8}=32 \ta_8\ ,\ {\cal
A}_{2,12}=48 \ta_{12}\ ,\ {\cal A}_{2,14}=56 \ta_{14}\ ,
\]
\[
{\cal A}_{2,18}=72 \ta_{18}\ ,\ {\cal A}_{2,20}=80 \ta_{20}\ ,\
{\cal A}_{2,24}=96 \ta_{24}\ ,\ {\cal A}_{2,30}=120 \ta_{30}\ ,
\]
\[
{\cal A}_{8,8}= \frac{21}{5} \ta_{14} +  \frac {7}{25} \ta_2
\ta_{12}\ ,\ {\cal A}_{8, 12}= - \frac{27}{7} \ta_2^5 \ta_8 +
\frac{2421}{5} \ta_2^2 \ta_{14} + 81 \ta_{18} -  \frac{14664}{7}
\ta_2 \ta_8^2
 -  \frac{12}{5} \ta_2^3 \ta_{12}\ ,
\]
\[
{\cal A}_{8,14}=75 \ta_{20} +  \frac{32}{15} \ta_8 \ta_{12} +
\frac{2 }{5} \ta_2 \ta_{18} \ ,\ {\cal A}_{8,18}= \frac{2349}{2}
\ta_2^2 \ta_{20} + \frac{16640}{21} \ta_8^3- 18 \ta_2^5 \ta_{14}
\]
\[
+ 270 \ta_{24} + \frac{3744}{49} \ta_2^4 \ta_8^2 +  \frac{88}{35}
\ta_2^2 \ta_8 \ta_{12}+ \frac{16}{75}\ta_{12}^2 - \frac{12232}{35}
\ta_2 \ta_8 \ta_{14} - \frac{27}{5} \ta_2^3 \ta_{18}\ ,
\]
\[
{\cal A}_{8,20}= \frac{3}{5} \ta_2 \ta_{24} + \frac{32}{375}
\ta_{12} \ta_{14}\ ,\ {\cal A}_{8,24}=\ta_{30} -
\frac{43264}{6615} \ta_2^3 \ta_8^3 +  \frac {2288}{525} \ta_2^4
\ta_8 \ta_{14} + \frac{16}{1125} \ta_2^2 \ta_{12} \ta_{14}
\]
\[
+ \frac {832  \ta_2^2 \ta_8 \ta_{18}}{1575} + \frac{64 \ta_{12}
\ta_{18}}{ 1125}  - \frac {126 \ta_2^5 \ta_{20}}{5} - \frac {39
\ta_2^3 \ta_{24} }{5} - \frac {9936 \ta_2 \ta_8 \ta_{20}}{35} +
\frac {311168 \ta_8^{ 2} \ta_{14}}{19845} +  \frac {352 \ta_2
\ta_{14}^{2 }}{125}\ ,
\]
\[
{\cal A}_{8,30}= \frac {48}{5} \ta_{12} \ta_{24} -
\frac{1555840}{1323} \ta_8^2 \ta_{20} - \frac{27}{5} \ta_2^3
\ta_{30} + \frac{43264}{945} \ta_8 \ta_{14}^2 +
\frac{43264}{19845} \ta_2 \ta_8^2 \ta_{18}
\]
\[
+ \frac{104}{7} \ta_2^2 \ta_8 \ta_{24} - \frac{106496}{14175}
\ta_2 \ta_8 \ta_{12} \ta_{14} + \frac{692224}{59535} \ta_8^3
\ta_{12} +  \frac{656}{15} \ta_2^2 \ta_{12} \ta_{20} -
\frac{1016}{5} \ta_2 \ta_{14} \ta_{20}\ ,
\]
\[
{\cal A}_{12,12}= \frac{22275}{2} \ta_2 \ta_{20} +
\frac{1185120}{49} \ta_2^3 \ta_8^2 + 11880 \ta_8 \ta_{14} +
\frac{810}{49} \ta_2^7 \ta_8
\]
\[
- \frac{5400}{7} \ta_2 \ta_8  \ta_{12} - \frac{43011}{7} \ta_2^4
\ta_{14} + 9 \ta_2^5 \ta_{12} -  \frac{7155}{7} \ta_2^2 \ta_{18}\
,
\]
\[
{\cal A}_{12,14}= \frac{164250}{7} \ta_2^2 \ta_{20} + 2560 \ta_8^3
+ 4050 \ta_{24} - \frac{240}{7} \ta_2^4 \ta_8^2 -  \frac{144}{7}
\ta_2^2 \ta_8 \ta_{12} - \frac{22752}{7} \ta_2 \ta_8 \ta_{14} -
\frac{6}{7} \ta_2^3 \ta_{18}\ ,
\]
\[
{\cal A}_{12,18}= -  \frac{2688}{5} \ta_2 \ta_{12} \ta_{14} +
12600 \ta_8 \ta_{20} - \frac{4415680}{49} \ta_2^2 \ta_8^3 -
\frac{2218185}{14} \ta_2^4 \ta_{20}+ 504 \ta_{14}^2
\]
\[
+ \frac{270}{7} \ta_2^7 \ta_{14} - \frac{72}{35} \ta_2^2
\ta_{12}^2 -  \frac {53640}{343} \ta_2^6 \ta_8^2 - \frac{264}{49}
\ta_2^4 \ta_8 \ta_{12} + \frac{1891584}{49} \ta_2^3 \ta_8 \ta_{14}
\]
\[
+ \frac{10112}{7} \ta_8^2 \ta_{12} + \frac{20040}{7} \ta_2 \ta_8
\ta_{18} + \frac{81}{7} \ta_2^5 \ta_{18} - \frac{381375}{14}
\ta_2^2 \ta_{24}\ ,
\]
\[
{\cal A}_{12,20} = 45 \ta_{30} -  \frac{48}{35} \ta_2^4 \ta_8
\ta_{14} - \frac{144}{175} \ta_2^2 \ta_{12} \ta_{14} -
\frac{9}{7}\ta_2^3 \ta_{24} + \frac{4800}{7} \ta_2  \ta_8 \ta_{20}
- \frac {315008}{2205} \ta_8^2 \ta_{14}
\]
\[
+ \frac {384}{25} \ta_2 \ta_{14}^2\ ,
\]
\[
{\cal A}_{12,24} = 54 \ta_2^7 \ta_{20} + 408 \ta_{14} \ta_{20} +
\frac{10112}{35} \ta_2^3 \ta_{14}^2 +  \frac{43264}{3087} \ta_2^5
\ta_8^3 + \frac{1024}{15} \ta_8^2 \ta_{18} +  \frac {117}{7}
\ta_2^5 \ta_{24}
 \]
 \[
- \frac{1504}{735} \ta_2^4 {\ta_8} \ta_{18} +
\frac{21408}{49}\ta_2^3 \ta_8 \ta_{20} - 576 \ta_2  \ta_{12}
\ta_{20} + \frac{11152}{175} \ta_2 \ta_{14} \ta_{18} +
\frac{32}{525}\ta_2^4 \ta_{12} \ta_{14}- \frac{64}{7} \ta_2^6
\ta_8 \ta_{14}
\]
\[
- \frac{60764416}{46305} \ta_2^2 \ta_8^2 \ta_{14} +
\frac{3328}{105}\ta_8 \ta_{12} \ta_{14} - \frac {96}{175} \ta_2^2
\ta_{12} \ta_{18} + \frac{6600}{7} \ta_2 \ta_8 \ta_{24} -
\frac{1968}{7} \ta_2^2 \ta_{30}\ ,
\]
\[
{\cal A}_{12,30}= -  \frac {3056}{7} \ta_2^4 \ta_{12} \ta_{20} -
\frac{36000}{49} \ta_2^6 \ta_8 \ta_{20} + 27000 \ta_{20}^2 -
\frac{1730560}{9261} \ta_2^4 \ta_8^4 + \frac{131456}{1029} \ta_2^5
\ta_8^2 \ta_{14}
\]
\[
-  \frac {346112 \ta_2^2 \ta_8^3 \ta_{12}}{3087} -  \frac{43264
\ta_2^3 \ta_8^2 \ta_{18}}{9261} -  \frac {9120 \ta_2^4 \ta_8
\ta_{24}}{49} -  \frac{648}{7} \ta_2^2 \ta_{12} \ta_{24} +
\frac{71552}{945} \ta_2^3 \ta_8 \ta_{12} \ta_{14}
\]
\[
-  \frac{31800}{7} \ta_2 \ta_8 \ta_{30}  + 1728 \ta_2 \ta_{14}
\ta_{24} +  \frac{81}{7} \ta_2^5 \ta_{30} +
\frac{55377920}{3969}\ta_8^5 - \frac{16640}{7} \ta_8 \ta_{12}
\ta_{20}
\]
\[
+ \frac {469779200}{3087} \ta_2^2 \ta_8^2 \ta_{20} -
\frac{95352}{7}\ta_2^3 \ta_{14} \ta_{20} -  \frac {28080}{7} \ta_2
\ta_{18} \ta_{20} + \frac {7444736}{2205} \ta_2^2 \ta_8 \ta_{14}^2
\]
\[
+ \frac{1646080}{49} \ta_8^2 \ta_{24} - \frac{263644160}{9261}
\ta_2 \ta_8^3 \ta_{14} + \frac {43264}{49} \ta_8 \ta_{14} \ta_{18}
\ ,
\]
\[
{\cal A}_{14,14} = - 1575 \ta_2^3 \ta_{20} -  \frac{525}{2} \ta_2
\ta_{24} - \frac{99200}{63} \ta_2 \ta_8^3 +  \frac{11360
}{21}\ta_2^2 \ta_8 \ta_{14} - \frac{8}{3} \ta_{12} \ta_{14} +
\frac{208}{3} \ta_8 \ta_{18}\  ,
\]
\[
{\cal A}_{14,18}= \frac{75}{2} \ta_{30} +  \frac{4421120 \ta_2^3
\ta_8^3}{441} +  \frac {1408 \ta_2 \ta_8^2 \ta_{12}}{21} - \frac
{23680 \ta_2^4 \ta_8 \ta_{14}}{7} + \frac{80 \ta_2^2 \ta_{12}
\ta_{14}}{3}-  \frac{2928}{7} \ta_2^2 \ta_8 \ta_{18}
\]
\[
+  \frac{32}{15} \ta_{12} \ta_{18} + 9450 \ta_2^5  \ta_{20} + 1575
\ta_2^3 \ta_{24} +  \frac{35080}{7} \ta_2 \ta_8 \ta_{20} +
\frac{1902848}{1323} \ta_8^2 \ta_{14} - 704 \ta_2 \ta_{14}^2\ ,
\]
\[
{\cal A}_{14,20} = 32 \ta_8 \ta_{24} -  \frac{16}{3} \ta_{12}
\ta_{20} +  \frac{32}{3} \ta_2^2 \ta_{14}^2 -  \frac{61696}{1323}
\ta_2 \ta_8^2 \ta_{14} +  \frac{48}{25} \ta_{14} \ta_{18} - 3
\ta_2 \ta_{30}\ ,
\]
\[
{\cal A}_{14,24}= -  \frac{16}{3} \ta_{12} \ta_{24} +
\frac{512}{9} \ta_8^2 \ta_{20} + 18 \ta_2^3  \ta_{30} +
\frac{2944}{75} \ta_8 \ta_{14}^2 + \frac{1856}{525} \ta_2^2
\ta_{14} \ta_{18} - \frac{64}{3} \ta_2^4 \ta_{14}^2 +
\frac{32}{25} \ta_{18}^2
\]
\[
- \frac{4352}{315} \ta_2 \ta_8^2 \ta_{18} +  \frac{2154496}{6615}
\ta_2^3 \ta_8^2 \ta_{14} -  \frac{9328}{7} \ta_2^4 \ta_8 \ta_{20}
- \frac{8696}{21} \ta_2^{ 2} \ta_8 \ta_{24}
\]
\[
+ \frac{1664}{525} \ta_2 \ta_8 \ta_{12} \ta_{14} -
\frac{1384448}{3969} \ta_2^2 \ta_8^4 - \frac{4544}{7} \ta_2
\ta_{14} \ta_{20} \ ,
\]
\[
{\cal A}_{14,30}= -  \frac{1664}{7} \ta_2 \ta_8 \ta_{12} \ta_{20}
- \frac{4630912}{19845} \ta_2 \ta_8 \ta_{14} \ta_{18} -
\frac{38600}{7} \ta_2 \ta_{20}^2 - \frac{2145894400}{250047} \ta_2
\ta_8^5 - 8 \ta_{12} \ta_{30}
\]
\[
+ \frac{26800}{21} \ta_2^2 \ta_{18} \ta_{20} + 1200 \ta_2^2
\ta_{14} \ta_{24} +  \frac {584 \ta_2^2 \ta_8 \ta_{30}}{7}-  \frac
{1459840}{189} \ta_2 \ta_8^2 \ta_{24}+  \frac{726968320}{83349}
\ta_2^2 \ta_8^3 \ta_{14}
\]
\[
+  \frac{346112}{6615} \ta_8^2 \ta_{12} \ta_{14} +
\frac{4499456}{11907} \ta_8^3  \ta_{18} + 216 \ta_{18} \ta_{24} -
\frac{2248832}{1323} \ta_8 \ta_{14} \ta_{20} -
\frac{17667200}{441} \ta_2^3 \ta_8^2 \ta_{20}
\]
\[
- \frac{83200}{63} \ta_2^3 \ta_8 \ta_{14}^2 + 7200 \ta_2^4
\ta_{14} \ta_{20}\ ,
\]
\[
{\cal A}_{18,18}= - 56700 \ta_2^7  \ta_{20} + 1020 \ta_{14}
\ta_{20} - 960 \ta_2^3 \ta_{14}^2 -  \frac{19125440}{343} \ta_2^5
\ta_8^3 + \frac{70400}{21} \ta_8^2 \ta_{18}
\]
\[
- 9450 \ta_2^5 \ta_{24} + \frac{15504}{7}\ta_2^4 \ta_8  \ta_{18} +
\frac{24064}{147} \ta_2^3 \ta_8^2 \ta_{12} + \frac{559980}{7}
\ta_2^3 \ta_8 \ta_{20} + 68 \ta_2 \ta_{12} \ta_{20}- 1760 \ta_2
\ta_{14} \ta_{18}
\]
\[
- 256 \ta_2^4 \ta_{12} \ta_{14} +  \frac{134880}{7} \ta_2^6 \ta_8
\ta_{14} - \frac{114368}{147} \ta_2^2 \ta_8^2 \ta_{14} +
\frac{1088}{15} \ta_8 \ta_{12} \ta_{14} -  \frac{464}{15} \ta_2^2
\ta_{12} \ta_{18}
\]
\[
+ \frac{2368}{105} \ta_2 \ta_8 \ta_{12}^2 + \frac{164100}{7} \ta_2
\ta_8 \ta_{24} + \frac{166400}{441} \ta_2 \ta_8^4 - \frac{945}{2}
\ta_2^2 \ta_{30} \ ,
\]
\[
{\cal A}_{18,20}= \frac{16}{5} \ta_{12} \ta_{24} +
\frac{10880}{7}\ta_8^2 \ta_{20} + 18 \ta_2^3 \ta_{30} +
\frac{1408}{75} \ta_8 \ta_{14}^2 - \frac{944}{75} \ta_2^2 \ta_{14}
\ta_{18}+ \frac{667904}{2205} \ta_2^3 \ta_8^2 \ta_{14}
\]
\[
- \frac{352}{5} \ta_2^4 \ta_{14}^2 + \frac{240}{7} \ta_2^4 \ta_8
\ta_{20} - \frac{1088}{7} \ta_2^2 \ta_8 \ta_{24} +
\frac{1408}{525} \ta_2 \ta_8 \ta_{12} \ta_{14}  + 48 \ta_2^2
\ta_{12} \ta_{20} -  \frac{5032}{5} \ta_2 \ta_{14} \ta_{20}\ ,
\]
\[
{\cal A}_{18,24}= \frac {3328}{23625} \ta_2 \ta_{12}^2 \ta_{14} -
\frac{7264}{7} \ta_2 \ta_{14} \ta_{24} + \frac{136}{15} \ta_2^2
\ta_{12} \ta_{24} - \frac{12784}{15} \ta_2 \ta_{18} \ta_{20} -
\frac{26172032}{15435} \ta_2^5 \ta_8^2 \ta_{14}
\]
\[
- \frac{26850304}{83349} \ta_2 \ta_8^3 \ta_{14} +
\frac{3048064}{441} \ta_2^2 \ta_8^2 \ta_{20} - \frac{4432}{175}
\ta_2^4 \ta_{14} \ta_{18} + \frac{45456}{7} \ta_2^6 \ta_8 \ta_{20}
+ \frac{647296}{11025} \ta_2^2 \ta_8 \ta_{14}^2
\]
\[
-  \frac {692224}{19845} \ta_2^2 \ta_8^3 \ta_{12} - \frac{592}{5}
\ta_2^4 \ta_{12} \ta_{20} + \frac{94592}{1575} \ta_8 \ta_{14}
\ta_{18} - \frac{8808}{35} \ta_2^3 \ta_{14} \ta_{20} +
\frac{287744}{2205} \ta_2^3 \ta_8^2 \ta_{18}
\]
\[
+  \frac{256}{45} \ta_8 \ta_{12} \ta_{20} + \frac{14344}{7}
\ta_2^{4 } \ta_8 \ta_{24} + \frac{1280}{21} \ta_2 \ta_8 \ta_{30} +
\frac{13568}{3675} \ta_2^3 \ta_8 \ta_{12} \ta_{14}
\]
\[
+ \frac{4352}{945} \ta_2 \ta_8  \ta_{12} \ta_{18} +
\frac{40960}{21} \ta_8^2 \ta_{24} - \frac{1888}{225} \ta_2^2
\ta_{18}^2  - 108 \ta_2^5 \ta_{30}
\]
\[
+  \frac{47071232}{27783} \ta_2^4  \ta_8^4 + \frac{256}{1125}
\ta_{12} \ta_{14}^2 + \frac{4512}{35} \ta_2^6 \ta_{14}^2  + 200
\ta_{20}^2\ ,
\]
\[
{\cal A}_{18,30}= - \frac{74054656}{46305} \ta_2 \ta_8^2
\ta_{14}^2 + \frac{743104}{147} \ta_8 \ta_{14} \ta_{24} -
\frac{5448}{7} \ta_2^4 \ta_8 \ta_{30} + \frac{692224}{297675}
\ta_8 \ta_{12}^2 \ta_{14}
\]
\[
- \frac{3328}{315} \ta_2 \ta_{12}^2 \ta_{20} +
\frac{30457856}{83349} \ta_2 \ta_8^4 \ta_{12} + \frac{737884160
}{27783} \ta_2 \ta_8^3 \ta_{20} - \frac{24960}{7} \ta_8 \ta_{18}
\ta_{20}
\]
\[
+ \frac{448640}{9} \ta_2^3 \ta_8^2 \ta_{24} +
\frac{260368000}{1029} \ta_2^5  \ta_8^2 \ta_{20} +
\frac{692224}{59535} \ta_8^2 \ta_{12} \ta_{18} -
\frac{21112832}{9261} \ta_2^2 \ta_8^3 \ta_{18}
\]
\[
- \frac{10771764224}{194481} \ta_2^4  \ta_8^3 \ta_{14} +
\frac{216}{5} \ta_2^2 \ta_{12} \ta_{30} - \frac{326400 \ta_2^6
\ta_{14} \ta_{20}}{7} + \frac{3757312}{441} \ta_2^5  \ta_8
\ta_{14}^2 - 7920  \ta_2^4 \ta_{14} \ta_{24}
\]
\[
- \frac{79740 \ta_2 \ta_{20} \ta_{24}}{7} - 1536 \ta_2 \ta_{14}
\ta_{30} - \frac{57600}{7} \ta_2^4 \ta_{18} \ta_{20} +
\frac{128}{5} \ta_{12} \ta_{14} \ta_{20} - 1416 \ta_2^2 \ta_{18}
\ta_{24}
\]
\[
+ \frac {294080 \ta_2^2 \ta_8 \ta_{14} \ta_{20}}{21} -  \frac
{229924864 \ta_2^2  \ta_8^2 \ta_{12} \ta_{14}}{416745} + \frac
{8000 \ta_2 \ta_8 \ta_{12} \ta_{24}}{21} +  \frac{3306368 \ta_2^3
\ta_8 \ta_{14} \ta_{18}}{2205}
\]
\[
+  \frac{422528 \ta_2^3 \ta_8 \ta_{12} \ta_{20}}{147} +  \frac
{95637667840 \ta_2^3 \ta_8^5}{1750329} -  \frac{248400 \ta_2^3
\ta_{20}^2}{ 7} +  \frac{19793453056 \ta_8^4  \ta_{14}}{5250987} +
\frac {3354880 \ta_8^2 \ta_{30}}{1323}\ ,
\]
\[
{\cal A}_{20,20}= -  \frac {4096 \ta_2 \ta_8 \ta_{14}^2}{525} -
\frac {32 \ta_8 \ta_{30}}{15} + \frac {692224 \ta_8^3 \ta_{14}}{
59535} + \frac{304 \ta_{14} \ta_{24} }{25} +  \frac {224 \ta_2^2
\ta_{14} \ta_{20}}{5} -  \frac{64 \ta_{18} \ta_{20}}{15}\ ,
\]
\[
{\cal A}_{20,24}= - \frac{1664}{63} \ta_2 \ta_8^2 \ta_{24} + \frac
{48}{25} \ta_{18} \ta_{24} + \frac{32}{3} \ta_2^2 \ta_8 \ta_{30} -
\frac{16}{225} \ta_{12} \ta_{30} + \frac{68864}{1575} \ta_8
\ta_{14} \ta_{20}
\]
\[
+  \frac{1792}{5625} \ta_{14}^3 - 976 \ta_2 \ta_{20}^2 +
\frac{512}{3375} \ta_2 \ta_8 \ta_{14} \ta_{18} +
\frac{379136}{7875}\ta_2^3 \ta_8 \ta_{14}^2 + \frac{6656}{118125}
\ta_2 \ta_{12} \ta_{14}^2
\]
\[
- \frac{21458944}{297675} \ta_2^2 \ta_8^3  \ta_{14} + \frac{346112
}{893025}\ta_8^2 \ta_{12} \ta_{14} - \frac{10944}{175} \ta_2^2
\ta_{14} \ta_{24} - \frac{5344}{25} \ta_2^4 \ta_{14} \ta_{20}
\]
\[
+  \frac {544}{15} \ta_2^2  \ta_{18} \ta_{20} - \frac{13312}{49}
\ta_2^3 \ta_8^2 \ta_{20} - \frac{6656}{945} \ta_2 \ta_8 \ta_{12}
\ta_{20}\ ,
\]
\[
{\cal A}_{20,30}= - \frac{32}{5} \ta_{18} \ta_{30} - \frac{1008
}{25}\ta_2^2 \ta_{14} \ta_{30} + \frac{327808}{1323} \ta_2 \ta_8^2
\ta_{30} + \frac{12193792}{11907} \ta_2^2  \ta_8^2 \ta_{14}^2 +
\frac{3461120}{441} \ta_2^2 \ta_8^3 \ta_{20}
\]
\[
- \frac{110080}{21} \ta_8 \ta_{20} ^2 + 30240 \ta_2^4 \ta_{20}^2 -
\frac{227072}{21} \ta_2^3  \ta_8 \ta_{14} \ta_{20} -
\frac{11318912}{6615} \ta_2 \ta_8 \ta_{14} \ta_{24} +
\frac{692224}{3969} \ta_8^3 \ta_{24}
\]
\[
+  \frac{1384448}{1488375} \ta_8 \ta_{12} \ta_{14}^2 -
\frac{346112}{11907} \ta_8^2 \ta_{12} \ta_{20} - \frac{6656}{1575}
\ta_2 \ta_{12} \ta_{14} \ta_{20} + 864 \ta_{24}^2 + \frac{896}{25}
\ta_{14}^2 \ta_{20}
\]
\[
+  \frac{4499456}{99225} \ta_8^2 \ta_{14} \ta_{18} -
\frac{6656}{63} \ta_2 \ta_8 \ta_{18} \ta_{20} - \frac{8893693952
}{5250987} \ta_2 \ta_8^4 \ta_{14} + \frac{71360}{7} \ta_2^2
\ta_{20} \ta_{24}\ ,
\]
\[
{\cal A}_{24,24}= - \frac{11208704}{2083725} \ta_2 \ta_8^2
\ta_{14}^2 + \frac{35456}{315} \ta_8 \ta_{14} \ta_{24} - \frac{256
}{5} \ta_2^4 \ta_8 \ta_{30} - \frac{36608}{354375} \ta_2^3
\ta_{12} \ta_{14}^2
\]
\[
- \frac{133120}{567} \ta_2 \ta_8^3 \ta_{20} +  \frac{5888}{675}
\ta_8 \ta_{18} \ta_{20} + \frac{247936}{1323} \ta_2^3 \ta_8^2
\ta_{24} + \frac{156416}{63} \ta_2^5 \ta_8^2 \ta_{20} -
\frac{15921152}{297675} \ta_2^2 \ta_8^3 \ta_{18}
\]
\[
+  \frac{59531264}{297675} \ta_2^4 \ta_8^3  \ta_{14} + \frac{64
}{75} \ta_2^2 \ta_{12} \ta_{30} + \frac{21184}{25} \ta_2 ^6
\ta_{14} \ta_{20} - \frac{36916736}{165375} \ta_2^5 \ta_8
\ta_{14}^2 - \frac{6952}{3} \ta_2 \ta_{20} \ta_{24}
\]
\[
+  \frac {154976}{525} \ta_2^4 \ta_{14} \ta_{24} + \frac{544
}{525} \ta_2 \ta_{14} \ta_{30} - \frac{22016}{75} \ta_2^4 \ta_{18}
\ta_{20} - \frac{128}{225}\ta_{12} \ta_{14} \ta_{20} +
\frac{126976}{70875} \ta_2 \ta_8 \ta_{18}^2
\]
\[
-  \frac{5632}{225} \ta_2^2 \ta_{18} \ta_{24} +  \frac{9707776
}{33075} \ta_2^2 \ta_8 \ta_{14} \ta_{20} - \frac{346112}{297675}
\ta_2^2 \ta_8^2 \ta_{12} \ta_{14} + \frac{26624}{354375} \ta_2
\ta_{12} \ta_{14} \ta_{18}
\]
\[
- \frac{6656}{945} \ta_2  \ta_8 \ta_{12} \ta_{24} + \frac{2077696
}{99225} \ta_2^3  \ta_8 \ta_{14} \ta_{18} + \frac{13312}{315}
\ta_2^3 \ta_8  \ta_{12} \ta_{20} + \frac{71991296}{250047} \ta_2^3
\ta_8^5
\]
\[
+  \frac {18688}{118125} \ta_2^2 \ta_{14}^3  - 1040 \ta_2^3
\ta_{20}^2 +  \frac{5537792}{535815} \ta_8^4 \ta_{14} -  \frac
{256}{135} \ta_8^2 \ta_{30} + \frac{5888}{16875} \ta_{14}^2
\ta_{18}\ ,
\]
\[
{\cal A}_{24,30} = \frac{337378902016}{26254935} \ta_2^3 \ta_8^4
\ta_{14} + \frac{928}{25} \ta_2^2 \ta_{18} \ta_{30} - \frac{23168
}{525} \ta_2^2 \ta_{14}^2 \ta_{20} - \frac{1517115392}{231525}
\ta_2^4 \ta_8^2 \ta_{14}^2
\]
\[
+ \frac{1280768}{19845} \ta_8 \ta_{14} \ta_{30} -
\frac{1731598336}{27783} \ta_2^4 \ta_8^3 \ta_{20} + \frac{384}{25}
\ta_{14} \ta_{18} \ta_{20} - \frac{27008}{7} \ta_8 \ta_{20}
\ta_{24} - \frac{31784}{21} \ta_2 \ta_{20} \ta_{30}
\]
\[
+ \frac{40368}{175}\ta_2^4 \ta_{14} \ta_{30} -
\frac{94142464}{1250235} \ta_2 \ta_8^4 \ta_{18} - \frac{373454848
}{83349}\ta_2^2 \ta_8^3 \ta_{24} - \frac{346112 \ta_8^2 \ta_{12}
\ta_{24}}{11907}
\]
\[
- \frac{68224}{49} \ta_2^3 \ta_8^2 \ta_{30} +
\frac{245120}{147}\ta_2^2 \ta_8 \ta_{20}^2 -
\frac{440816}{7}\ta_2^4 \ta_{20} \ta_{24} - \frac{4339712}{496125}
\ta_2 \ta_8 \ta_{14}^3 - 5352 \ta_2^2 \ta_{24}^2
\]
\[
- \frac{29948379136}{15752961} \ta_2^2 \ta_8^6 + \frac{692224
}{99225} \ta_8^2 \ta_{18}^2 + \frac{896}{25} \ta_{14}^2 \ta_{24} -
\frac{128}{3} \ta_{12} \ta_{20}^2 + \frac{36435902464}{393824025}
\ta_8^3 \ta_{14}^2
\]
\[
+ \frac{11075584}{35721} \ta_8^4 \ta_{20} - 186240 \ta_2^6
\ta_{20}^2 + \frac{149897984}{2205} \ta_2^5 \ta_8 \ta_{14}
\ta_{20} + \frac{71991296}{2679075} \ta_2 \ta_8^3 \ta_{12}
\ta_{14}
\]
\[
+ \frac{107614208}{416745} \ta_2  \ta_8^2 \ta_{14} \ta_{20} -
\frac{30457856}{4465125} \ta_2^2 \ta_8 \ta_{12} \ta_{14}^2 + \frac
{346112}{3969} \ta_2^2 \ta_8^2 \ta_{12} \ta_{20} + \frac{176384
}{4725} \ta_2^3 \ta_{12} \ta_{14} \ta_{20}
\]
\[
+ \frac{1288960}{1323} \ta_2^3 \ta_8 \ta_{18} \ta_{20} -
\frac{1814398976}{6251175} \ta_2^2 \ta_8^2 \ta_{14} \ta_{18} +
\frac {71777408}{6615} \ta_2^3 \ta_8 \ta_{14} \ta_{24} - \frac
{13312}{4725} \ta_2 \ta_{12} \ta_{18} \ta_{20}
\]
\[
+ \frac{3328}{525} \ta_2 \ta_{12} \ta_{14} \ta_{24} + \frac{6784
}{105} \ta_2 \ta_8  \ta_{18} \ta_{24} + \frac{2768896}{4465125}
\ta_8 \ta_{12} \ta_{14} \ta_{18} - \frac{1664}{945} \ta_2 \ta_8
\ta_{12} \ta_{30}\ ,
\]
\[
{\cal A}_{30,30}= - \frac{578880}{7} \ta_2 \ta_8 \ta_{24}^2 +
\frac{153485443072}{236294415} \ta_8^4 \ta_{12} \ta_{14} -
\frac{21139200}{7} \ta_2^5 \ta_8 \ta_{20}^2
\]
\[
- \frac{556201984}{27783} \ta_2^2  \ta_8^3 \ta_{30} -
\frac{346112}{3969} \ta_8^2 \ta_{12} \ta_{30} +
\frac{63870677811200 }{330812181} \ta_2^2 \ta_8^5 \ta_{14}+
\frac{1827200}{21} \ta_2^2 \ta_{14} \ta_{20}^2
\]
\[
+ \frac{346112}{147} \ta_8^2 \ta_{18} \ta_{24} -
\frac{57309224960}{750141} \ta_2 \ta_8^4 \ta_{24} - \frac{14400
}{7} \ta_2^4 \ta_{20} \ta_{30} - \frac{2043099136000}{1750329}
\ta_2^3 \ta_8^4 \ta_{20}
\]
\[
- 432 \ta_2^2 \ta_{24} \ta_{30} - \frac{96640}{7} \ta_8 \ta_{20}
\ta_{30} - \frac{316160}{63} \ta_2 ^3 \ta_{12} \ta_{20}^2 + 16704
\ta_{14} \ta_{20} \ta_{24} - \frac{3258990592}{35721} \ta_2^3
\ta_8^3 \ta_{14}^2
\]
\[
+ \frac{116396081152}{5250987} \ta_8^3 \ta_{14} \ta_{20} + \frac
{1307238400}{27783} \ta_2 \ta_8^2 \ta_{20}^2 +
\frac{119062528}{59535} \ta_2^2 \ta_8 \ta_{12} \ta_{14} \ta_{20} +
\frac{3441737728}{3750705} \ta_8^2 \ta_{14}^3
\]
\[
- \frac{472478875648}{78764805} \ta_2  \ta_8^3 \ta_{14} \ta_{18} -
\frac{1583808512}{8037225} \ta_2 \ta_8^2 \ta_{12} \ta_{14}^2 -
\frac{71991296}{27783} \ta_2 \ta_8^3 \ta_{12} \ta_{20}
\]
\[
+ \frac{4230234112}{27783} \ta_2^2 \ta_8^2 \ta_{14} \ta_{24} +
\frac{2903879680}{83349} \ta_2^2 \ta_8^2 \ta_{18} \ta_{20} +
\frac{3058432}{735} \ta_2^3 \ta_8  \ta_{14} \ta_{30}
\]
\[
- \frac{147142400}{147} \ta_2^3  \ta_8 \ta_{20} \ta_{24} +
\frac{1384448}{6615} \ta_8 \ta_{12} \ta_{14} \ta_{24} +
\frac{153620480}{147} \ta_2^4  \ta_8^2 \ta_{14} \ta_{20}
\]
\[
+ \frac{13312}{21} \ta_2 \ta_8 \ta_{18} \ta_{30} -
\frac{452727808}{19845} \ta_2 \ta_8 \ta_{14}^2 \ta_{20} -
\frac{6656}{7} \ta_2 \ta_{12} \ta_{20} \ta_{24}
\]
\begin{equation}
\label{e6.14}
 - \frac{46419987660800}{992436543} \ta_2 \ta_8^7 +
\frac{97332232192}{47258883} \ta_8^5 \ta_{18} - 3840 \ta_{18}
\ta_{20}^2 - \frac{384}{5} \ta_{14}^2 \ta_{30}\ ,
\end{equation}
and
\[
{\cal B}_{2}=- 4 \om  \ta_2 + 32 (1 + 30 \nu)\ ,\ {\cal B}_{8}= -
16 \om  \ta_8 - \frac{21}{10} (1+18\nu)  \ta_2^3\ ,
\]
\[
{\cal B}_{12}= - 24 \om  \ta_{12} -\frac{480}{7} (17+ 207 \nu )
\ta_2 \ta_8 + \frac{9}{2} (1 + 18\nu)  \ta_2^5\ ,
\]
\[
{\cal B}_{14}=- 28 \om  \ta_{14} - \frac{16}{15}(1+  30 \nu   )
\ta_{12} - \frac{8}{7}( 23+354\nu) \ta_2^2 \ta_8\ ,
\]
\[
{\cal B}_{18}= - 36 \om  \ta_{18} +\frac{12}{7} (25+282\nu)
\ta_2^4 \ta_8 + \frac{4}{5}(19 +198\nu)  \ta_2^2 \ta_{12}-
\frac{8}{5}(1037+7470\nu) \ta_2 \ta_{14}
\]
\[
+ \frac{1024}{21}(83 + 435\nu ) \ta_8^2\ ,
\]
\[
{\cal B}_{20}=  - 40 \om  \ta_{20} + \frac{512}{63}(2+39\nu) \ta_2
\ta_8^2 - \frac{376}{75}( 1+18\nu) \ta_2^2 \ta_{14} -
\frac{32}{75}(1+30\nu) \ta_{18}\ ,
\]
\[
{\cal B}_{24}= - 48 \om \ta_{24} + \frac{512}{4725}(23-195\nu)
\ta_2 \ta_8 \ta_{12} + \frac{256}{1575}( 859 + 3480\nu) \ta_8
\ta_{14}- \frac{16}{105}(15331 +76230\nu) \ta_2 \ta_{20}
\]
\[
+ \frac{16}{1575}(115 + 6678\nu) \ta_2^2 \ta_{18} -
\frac{256}{735}(191+5590\nu) \ta_2^3  \ta_8^2 +
\frac{8}{75}(161+4338\nu) \ta_2^4 \ta_{14}\ ,
\]
\[
 {\cal B}_{30}= - 60 \om  \ta_{30} -  \frac {76544
\ta_2 \ta_{12} \ta_{14}}{14175} + \frac{64}{945} (8935+74646\nu )
\ta_2^3 \ta_8 \ta_{14}+ \frac{256}{19845}(3637+49140\nu) \ta_2
\ta_8  \ta_{18}
\]
\[
- \frac{128}{1323} (128693+531090\nu) \ta_8 \ta_{20} +
\frac{346112}{59535} (7 - 30\nu)  \ta_8^2 \ta_{12} -
\frac{4}{7}(839+2646\nu) \ta_2^2 \ta_{24}
\]
\begin{equation}
\label{e6.15}
 - \frac {13312}{27783} ( 2861 + 37362\nu  ) \ta_2^2
\ta_8^3 -16 (151 + 450 \nu ) \ta_2^4 \ta_{20} + \frac{64}{4725}(
5837 -  5670 \nu ) \ta_{14}^2\ .
\end{equation}

It can be found one-parametric algebra of differential operators
(in eight variables) for which ${\cal P}^{(1,3,5,5,7,7,9,11)}_n$
(see (\ref{e6.10f})) is a finite-dimensional irreducible
representation space. Furthermore, the finite-dimensional
representation spaces appear for different integer values of the
algebra parameter. They form an infinite non-classical flag which
coincides to ${\cal P}^{(E_8)}$ (\ref{e6.10f}). We call this
algebra $e^{(8)}$. Like the algebras $g^{(2)}, f^{(4)}$ introduced
in \cite{Rosenbaum:1998} and \cite{blt}, respectively,  in
relation to the $G_2$ and $F_4$ models, the algebra $e^{(8)}$ is
infinite-dimensional but finitely-generated. It will be described
elsewhere. The rational $E_8$ Hamiltonian in the algebraic form
(\ref{e6.13}) with coefficients (\ref{e6.14}) , (\ref{e6.15}) can
be rewritten in terms of the generators of this algebra.

The operator (\ref{e6.13}) is triangular in the basis of monomials
\linebreak 
$\ta_2^{p_2} \ta_8^{p_8} \ta_{12}^{p_{12}} \ta_{14}^{p_{14}}
\ta_{18}^{p_{18}} \ta_{20}^{p_{20}} \ta_{24}^{p_{24}}
\ta_{30}^{p_{30}}$. One can easily find the spectrum of (\ref{e6.13}),
\linebreak
$\nobreak{h_{\rm E_8}^{(r)} \varphi = -2\ep\varphi}$, explicitly
\begin{equation}
\label{e6.16}
  \ep_{n_1,n_2,n_3,n_4,n_5,n_6,n_7,n_8}= 2\, \om (n_1 + 4 n_2 + 6 n_3+
  7 n_4 + 9n_5 + 10n_6 + 12 n_7 + 15 n_8) \ ,
\end{equation}
where $p_i$ are non-negative integers. Degeneracy of the spectrum
is related to the number of partitions of integer number
$n=0,1,2,\ldots$ to $n_1 + 4 n_2 + 6 n_3+ 7 n_4 + 9n_5 + 10n_6 +
12 n_7 + 15 n_8$. The spectrum does not depend on the coupling
constants $g$, it is equidistant and corresponds to the spectrum
of a harmonic oscillator. Finally, the energies of the original
rational $E_8$ Hamiltonian (\ref{e6.1}) are $E=E_0+\ep$. As an
illustration the first eigenfunctions are presented in the
Appendix D. 

It is worth noting that the Hamiltonian (\ref{e6.13}) has a remarkable
property: there exists a family of eigenfunctions which depend on the
single variable $\ta_2$. These eigenfunctions are the associated
Laguerre polynomials. This property admits to construct a
quasi-exactly-solvable generalization of the rational $E_8$ model. It
will be done elsewhere. Due to enormous technical difficulties we were
unable to find explicitly the boundary of the configuration space (in
other words, the boundaries of the $E_{8}$ Weyl chamber) in the
Weyl-invariant variables $\ta$'s similar to what was previously done
for $G_2$ and $F_4$ cases (see Section 3 and 4, correspondingly).


\section{Conclusion}

\indent

We have found in uniform way that all the rational integrable models
associated with the root systems of exceptional Lie algebras are
exactly-solvable, thus continuing the analysis of the rational (and
trigonometric) systems related to the $A_n$ and $BC_n$ root systems as
well as their supersymmetric generalizations carried out in
Ref.\cite{Ruhl:1995, Brink:1997, Ruehl:1998, Ruehl:1999}.  Our method
contains two important ingredients: (i) gauging away the ground state
eigenfunction and (ii) taking specific Weyl-invariant polynomials
(functions) as variables. After this procedure each Hamiltonian takes
an algebraic form thus becoming a linear differential operator with
polynomial coefficients.  Furthermore, it has infinitely-many
invariant subspaces of polynomials, which form a {\it minimal} flag.
The known characteristic vectors of the minimal flags are collected in
the Table. The meaning of the Weyl-invariant variables (\ref{var1}),
(\ref{var2}), (\ref{e3.12var1}), (\ref{e3.15var2}), (\ref{e4.9}),
(\ref{e5.9}), (\ref{e6.10}) in which the flag of invariant subspaces
is minimal is unclear so far. Namely, why in these Weyl-invariant
variables the minimal flag is preserved. We show that unlike the
rational $A_n$ and $BC_n$ models which all are of the hypergeometric
type every rational $G_2, F_4, E_{6,7,8}$ model is special. Each of
them is characterized by its own hidden infinite-dimensional algebra
which deserve a separate investigation.  It will be done elsewhere.

Algebraic forms of the rational $G_2, F_4, E_{6,7,8}$ models which we
found allow to construct quasi-exactly-solvable generalizations
\cite{Turbiner:1988} of these models similar to those of the paper
\cite{Minzoni:1996}. It is already done in \cite{Turbiner:2004}. It
concludes an analysis of rational models associated with classical
(crystallographic) root systems, while the analysis of the rational
systems related with dihedral (non-crystallographic) root systems
$H_3, H_4, I_2(m)$ is still waiting to be done.

The presented work complements previous studies where the algebraic and
the Lie-algebraic forms as well as the corresponding flags were
found for the rational and trigonometric Olshanetsky-Perelomov
Hamiltonians of $A-D$ series (and their supersymmetric
generalizations) \cite{Ruhl:1995, Brink:1997}, $G_2$
\cite{Rosenbaum:1998} and $F_4$ \cite{blt} models. In order to
conclude a study of the whole set of Olshanetsky-Perelomov
integrable systems appearing in the Hamiltonian reduction method
it is necessary to perform the same analysis for remaining
$E_{6,7,8}$ integrable trigonometric models. We consider it as a
challenging task for future.

\small{
\section*{Acknowledgement}

The authors want to express a deep gratitude to N.~Nekrasov for
careful reading of the text, many valuable remarks and the interest to
the work.  K.G.B. thanks Instituto de Ciencias Nucleares, UNAM for
kind hospitality extended to him where the collaboration was
initiated. The work was supported in part by DGAPA grants No. {\it
  IN120199, IN124202}, CONACyT grant {\it 25427-E}, INTAS grant {\it
  00-00366}, CRDF RUP2-2621-MO-04, and by Russian Funds of Basic
Research 00-15-96786 and 03-02-04004, 04-02-17263 and SSh-1774-2003.2.
}

\begin{table*}
\begin{center}
\begin{tabular}{|c|c|c|}
\hline
             &           &               \\
\hspace{10pt} Model \hspace{10pt} & Rational  & \hspace{5pt}
Trigonometric \hspace{5pt}\\
             &           &               \\  \hline \hline
             &           &               \\
\raise 12pt\hbox{$A_n$} & $\buildrel\underbrace{(1,1,\ldots
1)}\over n$ &
$\buildrel\underbrace{(1,1,\ldots 1)}\over n$  \\
             &           &               \\ \hline
             &           &               \\
\raise 12pt\hbox{$BC_n$} & $\buildrel\underbrace{(1,1,\ldots
1)}\over n$ &
$\buildrel\underbrace{(1,1,\ldots 1)}\over n$ \\
             &           &               \\ \hline
             &           &               \\
$G_2$ & (1,2) & (1,2) \\
             &           &               \\ \hline
             &           &               \\
$F_4$ & (1,2,2,3) & (1,2,2,3) \\
             &           &               \\ \hline
             &           &               \\
$E_6$ & (1,1,2,2,2,3) & (1,1,2,2,2,3) \\
             &           &               \\ \hline
             &           &               \\
$E_7$ & (1,2,2,2,3,3,4) & ? \\
             &           &               \\ \hline
             &           &               \\
$E_8$ & (1,3,5,5,7,7,9,11) & ? \\
             &           &               \\ \hline
\end{tabular}
\caption{Characteristic vectors of rational and trigonometric
models associated with classical (crystallographic) root spaces}
\end{center}
\end{table*}

\clearpage


\appendix
\setcounter{equation}{0}

\setcounter{equation}{0}

\section{First eigenfunctions of the rational $F_4$ models}

In this Appendix we present explicit expressions for the first
eigenfunctions of the rational $F_4$ model at $n=0,1,2$.

\begin{alignat*}{3}
 &\bullet\ {\bf n=0}\ \mbox{(one eigenstate)} &\\
 &\quad & \phi_0\ =\ & 1\ ,\\
 &\quad & \ep_0\ =\ & 0 \ .  \\[8pt]
 &\bullet\ {\bf n=1}\ \mbox{(one eigenstate)}  &\\
 &\quad & \phi_1\ =\ &\  \ta_2 - \frac{2}{\om}(6\mu+6\nu+1)\ , \\
 &\quad & \ep_1\ =\ & 2\om \ . \\[8pt]
 &\bullet\ {\bf n=2} \ \mbox{(three eigenstates)} \\
 &\quad & \phi_2^{(1)}\ =\ &\  \ta_2^2\ -\ \frac{6}{\om}(4\mu+4\nu+1)
 \ta_2\ +\ \frac{6}{\om^2}(4\mu+4\nu+1)(6\mu+6\nu+1)\ , \\
 & & \ep_2^{(1)}\ =\ & 4\om\ , \\[5pt]
 &\quad & \phi_2^{(2)}\ =\ &  \ta_6\ -\ \frac{1}{4\om}(2\mu+4\nu+1)
\ta_2^2\ +\
\frac{3}{4\om^2}(2\mu+4\nu+1)(4\mu+4\nu+1) \ta_2\ \\
&&&
 + \frac{1}{2\om^3}(2\mu+4\nu+1)(6\mu+6\nu+1)(4\mu+4\nu+1)\ , \\
&\quad & \ep_2^{(2)}\ =\ & 6\om\ ,\\[5pt]
&\quad & \phi_2^{(3)}\ =\ &\  \ta_8 - \frac{1}{\om}(3\nu+1) \ta_6\
+\
\frac{1}{8\om^2}(3\nu+1)(2\mu+4\nu+1) \ta_2^2\ \\
&&&
 - \frac{1}{4\om^3}(3\nu+1)(2\mu+4\nu+1)(4\mu+4\nu+1) \ta_2\  \\
&&&
 +\frac{1}{8\om^4}(3\nu+1)(2\mu+4\nu+1)(6\mu+6\nu+1)(4\mu+4\nu+1)\ ,\\
 &\quad &\ep_2^{(3)}\ =\ & 8\om\ .
\end{alignat*}

\section{First eigenfunctions of the rational $E_6$ model}

\begin{itemize}
\item $n=0 \ \mbox{(one eigenstate)}$
\[
\phi_0= 1\ , \quad \ep_0=0\ ,
\]

\item $n=1 \ \mbox{(one eigenstate)}$

\[
\phi_{1,1} =  \ta_2  -  \frac{6}{\om} (1+12\nu)\ , \quad \ep_{1,1}
= 2 \om \ ,
\]

\[
\phi_{1,2} = \ta_5\ , \quad \ep_{1,2} = 5 \om \ .
\]

\item $n=2 \ \mbox{(six eigenstates)}$

\[
\phi_{2,1} = \ta_2^2 - \frac{16}{\om} (1+9\nu) \ta_2 +
\frac{48}{\om^2} (1+9\nu)(1+12\nu) \ , \quad \ep_{2,1} = 4\om\ ,
\]
\[
\phi_{2,2} = - 4\ta_6+ \frac{405}{\om} \ta_2^2 (1+6\nu)-
\frac{3240}{\om^2}\ta_2(1+6\nu)(1+9\nu)
+\frac{6480}{\om^3}(1+6\nu)(1+9\nu)(1+12\nu)\ ,
\]
\[
\ep_{2,2} = 6 \om \ ,
\]
\[
\phi_{2,3} = (\ta_2 -\frac{8}{\om}(2+9\nu))\ta_5\ , \quad
\ep_{2,3} = 7 \om\ ,
\]
\[
\phi_{2,4} = \ta_8-\frac{24}{\om} (1+3\nu)\ta_6 +
\frac{1215}{\om^2}( 1+3\nu)( 1+6\nu)\ta_2^{2} -
\frac{6480}{\om^3}( 1+3\nu)( 1+6\nu)( 1+9\nu)\ta_2
\]
\[
+\frac{9720}{\om^4}(1+3\nu)( 1+6\nu)( 1+9\nu)( 1+12\nu)\ , \quad
\ep_{2,4} = 8 \om\ ,
\]
\[
\phi_{2,5} = \ta_9 -\frac{54}{\om}(2+3\nu)\ta_2\ta_5
+\frac{216}{\om^2} (2+3\nu)(2+9\nu)\ta_5\ , \quad \ep_{2,5} =
9\om\ ,
\]
\[
\phi_{2,6} = \ta_5^2-\frac{2}{\om}\ta_8 + \frac{24}{\om^2}
(1+3\nu)\ta_6 - \frac{810}{\om^3} (1+3\nu)(1+6\nu)\ta_2^2
\]
\[
+\frac{3240}{\om^4} (1+3\nu)(1+6\nu)(1+9\nu)\ta_2 -
\frac{3888}{\om^5} (1+3\nu)(1+6\nu)(1+9\nu)(1+12\nu)\ , \quad
\ep_{2,6} = 10\om\ .
\]
\end{itemize}

\section{First eigenfunctions of the rational $E_7$ model}

\begin{itemize}

\item $n=0 \ \mbox{(one eigenstate)}$

\[
\phi_0  = 1  \ , \quad \ep_0=0\ .
\]

\item $n=1 \ \mbox{(one eigenstate)}$

\[
\phi_1  =   \ta_2  -  \frac {14}{\om}(1 + 18\nu )\ , \quad \ep_1=
2\om\ .
\]

\item $n=2 \ \mbox{(four eigenstate)}$

\[
\phi_{2,1}  =  \ta_2^2 -  \frac {36}{\om}(1+14\nu)\ta_2 +  \frac
{252}{\om^2}(1+14\nu) (1+18\nu)   \ , \quad \ep_{2,1}=  4\om\ ,
\]
\[
\phi_{2,2} = \ta_6 + \frac{6}{\om}(1 + 10\nu)\ta_2^2
-\frac{108}{\om^2}(1+10\nu)(1+14\nu)\ta_2 +
\frac{504}{\om^3}(1+10\nu )(1+14\nu)(1+18\nu)\ ,
\]
\[
\ep_{2,2}=  6\om\ ,
\]
\[
\phi_{2,3}  = \ta_8 - \frac{5}{\om }(1 + 6\nu )\ta_6 -
\frac{15}{\om^2}(1 + 6\nu)(1+10\nu)\ta_2^2 + \frac{180}{\om^3}(1 +
6\nu)(1+10\nu)(1+14\nu)\ta_2
\]
\[
- \frac{630}{\om^4}(1 + 6\nu)(1+10\nu)(1+14\nu)(1+18\nu)\ ,
\]
\[
\ep_{2,3}=  8\om\ ,
\]
\[
\phi_{2,4}  = \ta_{10} +  \frac{1}{2\om }(1+2\nu)\ta_8 - \frac
{5}{4\om^2}(1+2\nu)(1+6\nu)\ta_6 - \frac
{5}{2\om^3}(1+2\nu)(1+6\nu)(1+10\nu)\ta_2^2
\]
\[
+ \frac {45}{2\om^4}(1+2\nu)(1+6\nu)(1+10\nu)(1+14\nu)\ta_2
\]
\[
- \frac{63}{\om^5}(1+2\nu)(1+6\nu)(1+10\nu)(1+14\nu)(1+18\nu)\ ,
\]
\[
\ep_{2,4}=  10\om\ ,
\]
\end{itemize}

\section{First eigenfunctions of the rational $E_8$ model}

\begin{itemize}

\item $n=0 \ \mbox{(one eigenstate)}$

\[
\phi_{0}  = 1\ , \quad \ep_{0}=0\ .
\]

\item $n=1 \ \mbox{(one eigenstate)}$

\[
\phi_{1}  = \ta_2 - \frac {8}{\om}\,(1+30\nu) \ , \quad
\ep_{1}=2\,\om\ .
\]

\item $n=2 \ \mbox{(one eigenstate)}$

\[
\phi_{2}  = \ta_2^2 -  \frac {20}{\om }\,(1+24\nu)\,\ta_2
  + \frac {80}{\om^2}\,(1+24\nu)(1+30\nu)\ , \quad
\ep_{2}=4\,\om\ .
\]

\item $n=3 \ \mbox{(two eigenstates)}$

\[
\phi_{3,1}  = \ta_2^3 -
 \frac {36}{\om }\,(1+20\nu)\,\ta_2^2 +
 \frac {360}{\om^2}\,(1+20\nu)(1+24\nu)\,\ta_2  -
 \frac {960}{\om^3}\,(1+20\nu)(1+24\nu)(1+30\nu)\ ,
\]
\[
\ep_{3,1}=6\,\om\ ,
\]
\[
\phi_{3,2}  =  \ta_8 + \frac{21}{40\,\om}\,(1+18\nu)\,\ta_2^3   -
\frac {189}{20\,\om^2}\,(1+18\nu)(1+20\nu)\,\ta_2^2  +
\]
\[
\frac{63}{\om^3}\,(1+18\nu)(1+20\nu)(1+24\nu)\,\ta_2 -
\frac{126}{\om^{4}}\,(1+18\nu)(1+20\nu)(1+24\nu)(1+30\nu)\ ,
\]
\[
\ep_{3,2}=8\,\om\ .
\]

\item $n=4 \ \mbox{(two eigenstates)}$

\[
\phi_{4,1}  = \ta_2^4 - \frac {8}{\om}\,(7+120\nu)\,\ta_2^3  +
\frac {144}{\om^2}\,(1+20\nu)\,(7+120\nu)\,\ta_2^2  -
\]
\[
\frac{960}{\om^3}\,(1+20\nu)(1+24\nu)(7+120\nu)\,\ta_2   +
\frac{1920}{\om^4}\,(1+20\nu)(1+24\nu)(1+30\nu)(7+120\nu)\ ,
\]
\[
\ep_{4,1}=8\,\om\ ,
\]
\[
\phi_{4,2}  =  \ta_2\,\ta_8 + \frac
{21}{40\,\om}\,(1+18\nu)\,\ta_2^4 - \frac
{24}{\om}\,(1+10\nu)\,\ta_8  - \frac {21}{\om
^2}\,(1+15\nu)\,(1+18\nu)\,\ta_2^3  +
\]
\[
\frac {252}{\om^3}\,(1+15\nu)(1+18\nu)(1+20\nu)\,\ta_2^2 - \frac
{1260}{\om^4}\,(1+15\nu)(1+18\nu)(1+20\nu)(1+24\nu)\,\ta_2 +
\]
\[
\frac
{2016}{\om^5}\,(1+15\nu)(1+18\nu)(1+20\nu)(1+24\nu)(1+30\nu)\ ,
\]
\[
\ep_{4,2}=10\,\om\ .
\]

\end{itemize}

\newpage

\begingroup\raggedright
\endgroup

\end{document}